\documentclass[9pt,twocolumn,twoside]{pnas-new}

\templatetype{pnasresearcharticle} 

\title{Magnetized Topological Insulator Multilayers}

\author[a]{Chao Lei}
\author[a,b]{Shu Chen} 
\author[a,1]{Allan H. MacDonald}

\affil[a]{Department of Physics, The University of Texas at Austin, Austin, Texas 78712, USA}
\affil[b]{Department of Physics, Shanghai University, Shanghai 200444, China}

\leadauthor{MacDonald} 

\significancestatement{Topological insulators like the van der Waals crystal Bi$_2$Te$_3$
have Dirac-cone surface states protected by time-reversal symmetry (TRS).  
Weak magnetic order in topological insulator thin films can lead to the 
quantum anomalous Hall effect (QAHE), which was first realized by magnetic doping of 
topological insulators to establish a fragile ferromagnetic state.
Recent work has established MnBi$_2$Te$_4$ as a structurally ordered magnetic topological insulator. 
We report on a theoretical study of thin films formed from MnBi$_2$Te$_4$ and Bi$_2$Te$_3$ 
building blocks that employs a simplified model validated by {\it ab initio} density-functional-theory.
We use it to shed light on the dependence of the quantum anomalous Hall effect on 
film thickness, magnetic configuration, the stacking sequences of the van der Waals coupled  
building blocks, and bulk electronic structure.}

\authorcontributions{A.H.M.and C.L. designed the research; 
C.L. and S.C. performed the research, C. L. and A.H.M. analyzed the data and wrote the paper.}
\authordeclaration{The authors declare no conflict of interest.}
\correspondingauthor{\textsuperscript{1}To whom correspondence should be addressed. E-mail: macd@physics.utexas.edu}


\begin{abstract}
We discuss the magnetic and topological properties of bulk crystals and quasi-two-dimensional
thin films formed by stacking intrinsic magnetized topological insulator ( for example  Mn(Sb$_{x}$Bi$_{1-x}$)$_2$X$_4$ with X = Se,Te)
septuple layers and topological insulator quintuple layers in arbitrary order.
Our analysis makes use of a simplified 
model that retains only Dirac-cone degrees of freedom on both 
surfaces of each septuple or quintuple layer.  We demonstrate the model's applicability and 
estimate its parameters by comparing with {\it ab initio }
density-functional-theory(DFT) calculations.  We then
employ the coupled Dirac cone model to provide an explanation for the 
dependence of thin-film properties, particularly the presence or absence of the 
quantum anomalous Hall effect, on film thickness, magnetic configuration, and 
stacking arrangement, and to comment on the design of Weyl superlattices.
\end{abstract}

\dates{This manuscript was compiled on \today}
\doi{\url{www.pnas.org/cgi/doi/10.1073/pnas.XXXXXXXXXX}}

\newcommand{\bk}{{\mathbf k}}
\newcommand{\bs}{{\mathbf \sigma}}
\setboolean{displaywatermark}{false}
\usepackage{amsmath,mathtools}
\newcommand{\angstrom}{\text{\normalfont\AA}}
\usepackage{changes}
\usepackage{ifpdf}
\setboolean{displaywatermark}{false}

\begin{document}

\maketitle
\thispagestyle{firststyle}
\ifthenelse{\boolean{shortarticle}}{\ifthenelse{\boolean{singlecolumn}}{\abscontentformatted}{\abscontent}}{}



\dropcap{T}opological insulator (TI) thin films in which time-reversal symmetry is
broken by magnetic order\cite{Tokua2019_MTI_review} have long been recognized as a promising platform
for the interplay between transport and magnetic properties that powers
spintronics.  Indeed the quantum anomalous Hall (QAH) effect, a high point of 
topological spintronics characterized by dissipationless transport, was first observed\cite{Yu_2010,Chang2013} in
magnetic topological insulator (MTI) thin films with ferromagnetic order, and strong magneto-electric response properties are 
expected in antiferromagnetic TI films\cite{Wang2016,Li2010,Mong2010}.
MTIs were first produced simply by doping 
(Sb$_{x}$Bi$_{1-x}$)$_2$X$_3$ TI thin films with magnetic elements. 
However, disorder, thought to be due mainly to inhomogeneity of the magnetic dopants\cite{Lee2015}, 
leads to complex magnetic order in these systems.  As a consequence the QAH effect appears 
only at extremely low temperatures, for example only at $\sim 30$ mK in Cr-doped 
Bi$_2$Te$_3$ , even though the Curie temperature is $\sim 15$K\cite{Chang2013}.  
For this reason the recent identification\cite{Otrokov_2017} of the 
Mn(Sb$_{x}$Bi$_{1-x}$)$_2$X$_4$ family of layered van der Waals materials, which 
can be viewed as MTIs that have magnetic moments on an ordered lattice, is a promising advance.

Important progresses have been made
in understanding the bulk and epitaxial thin film properties 
of this family of materials, both theoretically\cite{Otrokov_2017,Eremeev2017,Otrokov2019,Zhang2019,Li2019_theory,Chowdhury_2019} 
and experimentally\cite{Lee2013,Rienks2019,Zeugner2019,Yan2019,Lee2019,Li2020,Otrokov2019_film,Liu2020,Chen2019_Pressure,Deng2020,Gong2019,Zhang2019_AHC,Li2019,Hao2019,Chen2019,Ge2020,Hu2020,Ding2020}.
The quantum anomalous Hall effect has now been observed
in the presence of a relatively weak magnetic fields $\sim 5$T for thicknesses
ranging from 3-10 septuple layers\cite{Deng2020,Ge2020,Liu2020},
and large (almost quantized) anomalous Hall effects have been observed in the absence of an 
external magnetic field in high-quality five-septuple-layer MnBi$_2$Te$_4$ films\cite{Deng2020},
all at temperatures exceeding $1$K.  
The ratio of the quantum anomalous Hall temperature to the 
magnetic ordering temperature, $\sim 20$K, is much higher\cite{Ge2020} than in doped MTI films.
Some \cite{Zeugner2019,Otrokov2019,Vidal2019b,Lee2019} (but not all \cite{Li2019,Hao2019,Chen2019,Swatek2020})
photoemission experiments have identified the large surface state gaps $\sim 100$meV that are  
generally expected \cite{Rienks2019,Otrokov2019} theoretically in MTIs.    

Mn(Sb$_{x}$Bi$_{1-x}$)$_2$X$_4$ is a layered material composed of  
seven-layers X-(B,Sb)-X-Mn-X-(Bi,Sb)-X units that are coupled by weak van der Waals interactions.
These septuple layers may be viewed as 
(Bi,Sb)$_2$X$_3$ quintuple layers
in which the middle $X$ layer is replaced by an X-Mn-X trilayer.  
In agreement with theoretical predictions\cite{Eremeev2017,Li2019_theory}, 
neutron scattering measurements \cite{Li2019,Yan2019,Hu2020} show that bulk MBX has 
A-type antiferromagnetic order with Mn ions ordered ferromagnetically within each
septuple layer and antiferromagnetically between adjacent 
septuple layers. 
Because the antiferromagnetic 
interactions between septuple layers are weak, 
the Mn layer moments in thin films can be aligned by 
magnetic fields $\sim 5$ Tesla.
Thin films can be obtained either by epitaxial growth or by mechanical exfoliation from bulk crystals
\cite{Hirahara_2017,Otrokov_2017,Hagmann2017,Hirahara_2017,Eremeev2018,Wu2019,Vidal2019,Klimovskikh2019,Ding2020,Hu2020}.
Magnetic fields establish quantum Hall effects in thin 
films not by establishing 
Landau quantization, but by changing the magnetic configuration from antiferromagnetic to ferromagnetic through overcoming the weak interlayer 
exchange interactions \cite{Deng2020,Ge2020,Liu2020}.  

In this paper, we develop a simple model that can be used to address the 
properties of thin films and bulk crystals formed by stacking Mn(Sb$_{x}$Bi$_{1-x}$)$_2$X$_4$ 
septuple layers and (Sb$_x$,Bi$_{1-x}$)$_2$X$_3$ quintuple layers in arbitrary order.
The model contains only Dirac cone degrees of 
freedom on each surface of each septuple or quintuple layer.  
By comparing with {\it ab initio} density-functional-theory (DFT) 
calculations we are able to establish that this highly simplified 
description is usually
accurate, and also to fix values of the model's material-dependent parameters.
The advantages of the simplified model are that it facilitates the 
descriptions of crystals with complex stacking sequences and thin-films containing 
many layers, and more importantly that it allows trends in 
magnetic and topological properties across the family of materials to be 
recognized and understood.  Important materials variations include 
changes in the Sb fraction $x$ on the pnictogen sites, substitutions of X=S,Se,Te on the different chalcogen sites,
and growth controlled defect concentrations. 
Our model can also be used as a starting point for theories that 
account for gating electric fields, external magnetic fields, disorder,
and other perturbations that are difficult to describe using \textit{ab initio} approaches. 

In this paper we focus on Bi$_2$X$_3$ and MnBi$_2$X$_4$ 
with X=Se,Te, which have received greatest attention to date.
We find that bulk ferromagnetic MnBi$_2$X$_4$ (MBX) 
is a nearly ideal Weyl semimetal, 
that thin-film ferromagnets are two-dimensional Chern insulators with
Chern numbers that grow and gap sizes that decline with film thickness, 
and that antiferromagnetic thin films with sufficiently large odd layer numbers are Chern insulators with Chern number $|C|=1$.
We further find that ferromagnetic [MnBi$_2$X$_4$]$_{M}$[Bi$_2$X$_3$]$_N$
superlattices formed by inserting non-magnetic quintuple layers 
in the stack are ordinary insulators for $M/N$ smaller than about 3, but
become Weyl semimetals for larger $M/N$. For superlattices with (M,N) = (1,1), there is a 
large chance to acheive a Weyl semimetal phase with MnBi$_2$Te$_4$ as magnetic topological insulator layers and 
Bi$_2$Se$_3$, Bi$_2$Sb$_3$, or Sb$_2$Te$_3$ as topological insulator layers.

\section*{Coupled Dirac Cone Model}
We construct a model for a magnetized topological insulator multilayer 
by including Dirac cone degrees of freedom not only on the surface layers\cite{Dirac_Cone_2009,Rosenberg2012}, 
but also on the top and bottom of each magnetic septuple layer 
and non-magnetic quintuple layer as 
illustrated in Fig.~\ref{superlattice}.
We allow for arbitrary stacking of magnetic and non-magnetic layers either to form a thin film, 
or if repeated to form a bulk crystal.  

\ifpdf
\begin{figure}[htp]
\includegraphics[width=0.9\linewidth]{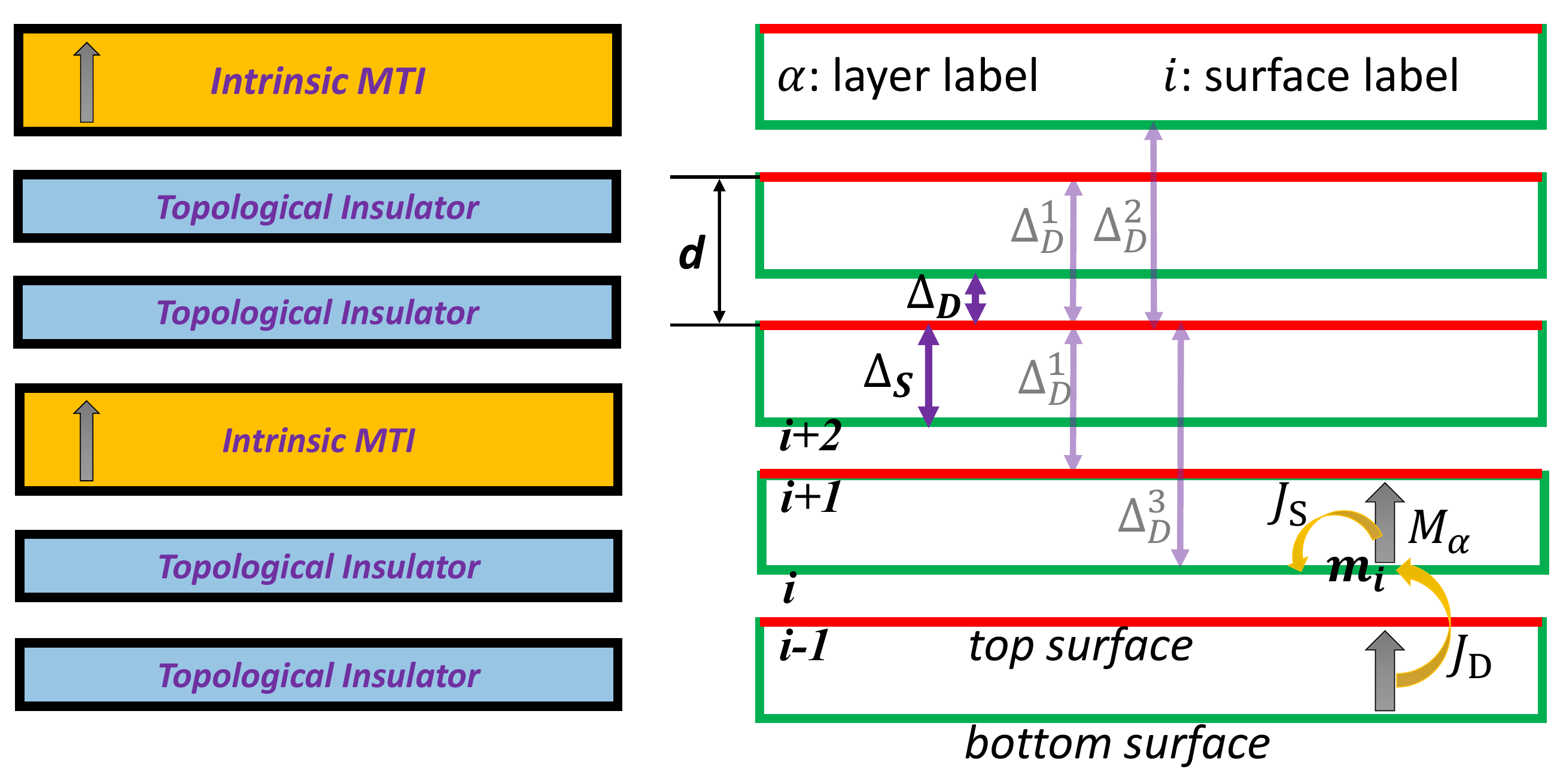}
\caption{\label{superlattice} Coupled Dirac Cone Model:
(Left Panel) Magnetized topological insulator multilayers composed of
M magnetic septuple layers and N non-magnetic quintuple layers.
(Right panel) Dirac cone model in which 
Dirac cones localized on the top and 
bottom of each layer are hybridized with remote Dirac-cones and altered by exchange interactions with the local moments present in the magnetic layers.}
\end{figure}
\fi
We allow for exchange interactions with the Mn local moments and arbitrary spin-independent hybridization
between different surfaces, denoting the 
hopping parameter between the i$^{th}$ surface and the j$^{th}$ surface by
$ \Delta_{ij}$.
The Hamiltonian is therefore:
\begin{equation}
\begin{split}
   H = & \sum_{\bk_{\perp},ij} \Big[\Big( \, 
   (-)^i  \hbar v_{_D}  (\hat{z} \times \bs) \cdot \bk_{\perp} + m_{i} \sigma_z \Big) \delta_{ij}   \\
   & + \Delta_{ij}(1-\delta_{ij} ) \Big] c_{\bk_{\perp} i}^{\dagger} c_{\bk_{\perp} j} ~,
\end{split}
\end{equation}
where spin-labels have been left implicit, $i$ and $j$ are Dirac cone labels with even integers reserved for layer bottoms and odd for layer tops, 
$ \hbar$ is the reduced Planck constant,
and $v_{_D}$ is the velocity of the Dirac cones.
The most important hybridization parameters, hopping within the same layer
($\Delta_S$) and hopping across the van der Waals gap between adjacent layers ($\Delta_D$),
are highlighted in Fig. \ref{superlattice}.
The mass gaps $m_i$ of the individual Dirac cones result from exchange interactions with 
Mn local moments and break time reversal symmetry: 
\begin{equation}
m_{i} = \sum_{\alpha} J_{i\alpha} M_{\alpha},
\end{equation}
where $ \alpha $ is a layer label.  
We limit our attention here to the case of magnetization perpendicular to the van der Waals layers 
so that $M_{\alpha} = \pm 1$ specifies the sense of magnetization on layer 
$\alpha$ if it is magnetic, and $M_{\alpha}=0$ on non-magnetic layers. 
Each Dirac cone can have two nearby Mn layers, 
one from the same layer with exchange splitting $J_S$ and  
one from the adjacent layer with exchange splitting $J_D$,
shown as in Fig. \ref{superlattice}), if these layers are magnetic.
In our surveys of MTI superlattice properties we retain only 
$\Delta_S$,$\Delta_D$, $J_{S}$ and $J_{D}$, which are normally dominant, 
as model parameters, the model allows different values of $\Delta_S$ in TI and MTI layers, and allows $\Delta_D$ at magnetic/non-magnetic heterojunctions to differ its value at magnetic/magnetic or non-magnetic/non-magnetic heterojunctions.

The model simplifies when only one type of layer is present.
When that layer is magnetic and the magnetic configuration is 
ferromagnetic, our model reduced to the topological insulator multilayer model 
proposed by Burkov and Balents \cite{Burkov_Balents_2011}. 
Instead of inserting a normal insulating layer between magnetic topological insulator layers to form a flexible three-dimensional system, as imagined 
in Ref.~\cite{Burkov_Balents_2011}, in this model the coupling between Dirac cones on different layers is  
across the van der Waals gap.   As we discuss below, some of the 
flexibility that would be associated with non-magnetic spacer layers of variable thickness,
can be recovered by inserting non-magnetic TI layers between magnetic TI layers.  

\section*{Comparison to DFT calculations} 
The relevance of the model that retains only Dirac-cone degrees of freedom to realistic systems can 
be tested by comparing with {\it ab initio} density-functional-theory (DFT) electronic structure calculations,
which we performed using a LDA+U approximation\cite{LDAU_Himmetoglu_2013}
with U = 5.25 eV on Mn sites. 

\ifpdf
\begin{figure}[htp]
\includegraphics[width=0.9\linewidth]{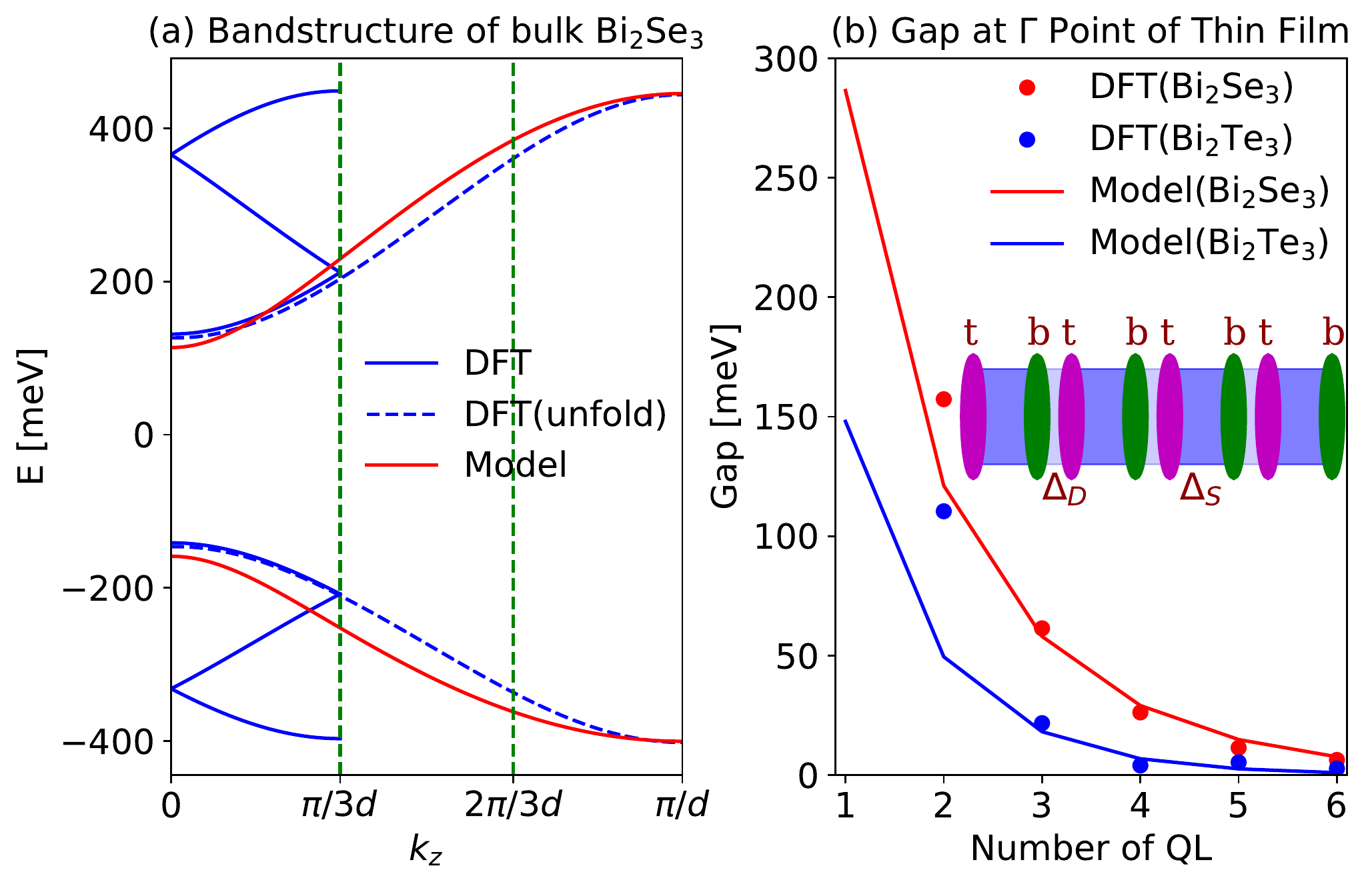}
\caption{\label{band_bs} Band structure of Bi$_2$Se$_3$ along the $\Gamma$ to $Z$ line
from DFT calculations and the Dirac cone model. (a) The blue line is the result from DFT calculations. Since the bulk Bi$_2$Se$_3$ layer stacking arrangement has three quintuple layers per unit cell, the DFT bands must be unfolded to the larger Brillouin zone. 
Fitting the bulk 
DFT band dispersion from $\Gamma$ to $Z$ yields $\Delta_S$ = 143 meV, $\Delta_D$ = -280 meV  and $\Delta_D^1$ = -11 meV.  The sign of $\Delta_S$ has been fixed by noting that  
the single-layer conduction band state at $\Gamma$ has greater weight in the 
middle of the layer than the valence band state.  The corresponding
Bi$_2$Te$_3$ fit yields $\Delta_S$ = 74 meV, $\Delta_D$ = -200 meV, 
and $\Delta_D^1$ = -10 meV. 
(b) Comparison between DFT and Dirac cone models 
for thin film gaps at the 
two-dimensional $\Gamma$ point {\it vs.} layer number. In the displayed scale, the gaps from DFT calculations are out of the displayed range,
it is around 847 meV for Bi$_2$Se$_3$ and around 488 meV for Bi$_2$Te$_3$.
The Dirac cone models were obtained by fitting to the bulk band dispersion. 
In the schematic illustration of the Dirac cone bonding network, 
t/b label the top/bottom surfaces of individual layers 
represented by magenta/green ellipses and the purple and grey links
represent hopping within and between layers.
}
\end{figure}
\fi

\subsection*{Non-Magnetic} 

As a first test of the model, we examine the bands of bulk Bi$_2$Se$_3$ as estimated by DFT.
We find that we can obtain a reasonable approximation to the DFT bands 
retaining only the hybridization parameters $\Delta_S$ for hopping within the same layer, 
and $\Delta_D$ for hopping to the adjacent surface of the adjacent layer.  The fit can
be improved by adding $\Delta_D^1$, for hopping to the same surface of the adjacent layer.  
Retaining only these three parameters 
the model predicts two spin-degenerate 
bands along the $\Gamma$ to $Z$ line in the Brillouin zone with dispersions
\begin{equation}
    E(k_z)  = \pm \sqrt{\Delta_S^2 + \Delta_D^2 + 2\Delta_S \Delta_D \cos k_zd} + 2\Delta_D^1 \cos k_zd,
\end{equation}
with d the distance between layer centers and $k_z$ the momentum along the $\Gamma$-$Z$ line. 
The corresponding DFT bands are illustrated in Fig.~\ref{band_bs}.
Because of details of the stacking arrangements 
that are not captured by the simplified model, the Bi$_2$Se$_3$ lattice used in the 
DFT calculations repeats only after three layers, instead of after a single layer.  
Nevertheless we find excellent agreement between the two
sets of bulk bands once the DFT bands are unfolded to the larger Brillouin-zone.
The property that the gap between the low energy bands is larger at $Z$ than at $\Gamma$ implies
that $\Delta_S$ and $\Delta_D$ have opposite signs. The parameter $\Delta_D^1$ is added to account for the 
small difference between the average of the DFT conduction and valence band energies at $\Gamma$ and $Z$.  
The model band energies at the $\Gamma$ point ($k_zd=0$)and the $Z$ point ($k_xd=\pi$) 
are respectively $ \pm |\Delta_S + \Delta_D| + 2\Delta_D^1 $ and $ \pm |\Delta_S - \Delta_D| - 2\Delta_D^1 $.
Fitting to these four energies we obtain $\Delta_S$ = 143 meV, $\Delta_D$ = -280 meV  and 
$\Delta_D^1$ = -11 meV.  The corresponding fitted parameters for Bi$_2$Te$_3$ are 
$\Delta_S$ = 74 meV, $\Delta_D$ = -200 meV, and $\Delta_D^1$ = -10 meV.  
The property that $|\Delta_D|$ is larger than $|\Delta_S|$, which is responsible for 
band inversion at $\Gamma$ and hence for the non-trivial band topology, is not surprising 
since the former hopping parameter is across a narrow van der Waals gap whereas the latter 
is across a wider quintuple layer.  As seen in Fig.~\ref{band_bs} (a) the unfolded DFT bands
are in excellent agreement with this simple model.

Using these parameters estimated from the bulk bands at $\Gamma$ and $Z$, 
we calculated the gap at the two-dimensional $\Gamma$ point for Bi$_2$Se$_3$ and Bi$_2$Te$_3$ 
thin films with thicknesses ranging from 1 to 6 quintuple layers.
As illustrated in Fig. \ref{band_bs} (b),
the gaps of thin films from the simplified model are in good agreement with DFT results (shown with red and blue dots), except in the single quintuple layer case for Bi$_2$Se$_3$ and in the single and double quintuple layer cases for Bi$_2$Te$_3$.
For the latter material the Dirac point lies in the valence bands instead of in the bandgap.  
This good agreement supports a physical picture in which the bulk gap is due to 
the hybridization of the Dirac cones network.  The thin film gap 
is small because the top(bottom) Dirac cone of the top(bottom) layer has no hybridization partner from adjacent layers (shown as inserted plot in Fig. \ref{band_bs} (b)).  
The thin film gap is due to hopping between surface layer Dirac cones via hybridized, and 
therefore gapped, Dirac cones in the interior.  
Because $\Delta_D^1$ is small compared to $\Delta_D$ we include 
only $\Delta_S$ and $\Delta_D$ in the following analysis.
The model's Dirac velocity parameter is 
estimated from the dependence of the DFT bands on 
in-plane momenta. (See Fig.S1 and Table S1 in the SI Appendix.)

\subsection*{Magnetic} 

We now turn to the magnetic case, starting with 
bulk MnBi$_2$Se$_4$(MBS) and MnBi$_2$Te$_4$(MBT). 
Below we refer to the magnetic configuration 
in which all magnetic layers are aligned as ferromagnetic.  
The ground state is antiferromagnetic, but because 
magnetic interactions between different van der Waals layers are extremely weak,
the ferromagnetic configuration can be realized by applying a relatively 
modest external magnetic field.
For the ferromagnetic configuration the exchange energies are the same in every layer and 
the model band energy dispersion
along the $\Gamma$ to $Z$ line is
\begin{equation}\label{en_mbs_fm}
 E(k_z)  = \pm \sqrt{\Delta_S^2 + \Delta_D^2 + 2\Delta_S \Delta_D \cos k_zd} \pm m_F ,
\end{equation}
where the ferromagnetic exchange splitting is $m_F=J_S+J_D$ is the sum of the on-layer ($J_S$) and 
neighboring-layer ($J_D$) exchange interactions.  
The energies at $k_zd=0$ and $k_zd=\pi$ are 
\begin{equation}\label{en_mbs_gz}
\begin{split}
    & E_{\Gamma} = \pm (\Delta_S + \Delta_D) \pm m_F \\
    & E_{Z} = \pm (\Delta_S - \Delta_D) \pm m_F.
\end{split}
\end{equation}
Because $m_F$ depends only on $J_S + J_D$ we must consider other magnetic configurations in
order to fit $J_S$ and $J_D$ independently.  For the 
antiferromagnetic configuration the exchange splitting 
$m_{AF} = J_S-J_D$ alternates in sign from layer to layer and 
the gap between conduction and valence bands at the $\Gamma$ point is
\begin{equation}
E_{gap}^{AF} = 2(\sqrt{\Delta_D^2+m_{AF}^2} -\Delta_S).
\end{equation}
By comparing these expressions with the DFT band dispersions of
bulk MnBi$_2$X$_4$(X = Se,Te), we extract the model
parameters summarized in Table \ref{para_table}.


\begin{table}\caption{Model parameters for MnBi$_2$X$_4$(X = Se,Te) in units of meV. 
For the column labelled DFT/EXP we use the experimentally available value of the 
antiferromagnetic state gap, and DFT calculation results for quantities that were 
not available from experiment.}\label{para_table}
\begin{tabular}{c|c|c|c|c}
 \hline
    Magnetic  & MBS & MBS & MBT & MBT \\
    Configuration  & (Model) & (DFT/Exp.) & (Model) & (DFT/Exp.) \\
 \hline
    $\Delta_S$  & 190  & -- & 84  &  -- \\
    $\Delta_D$  & -232 & -- & -127 & -- \\
    $J_S$       & 32   & -- & 36   & -- \\
    $J_D$       & 25   & -- & 29   & -- \\
    $E_{gap}^{\Gamma}$(AF) & 85 & $\simeq$100 & 86 & 50-200 \\
    $E_{gap}^{\Gamma}$(FM) & 30 & $\simeq$23 & 44 & 12 \\
 \hline
\end{tabular}
\end{table}

\ifpdf
\begin{figure}[htp]
\includegraphics[width=0.9\linewidth]{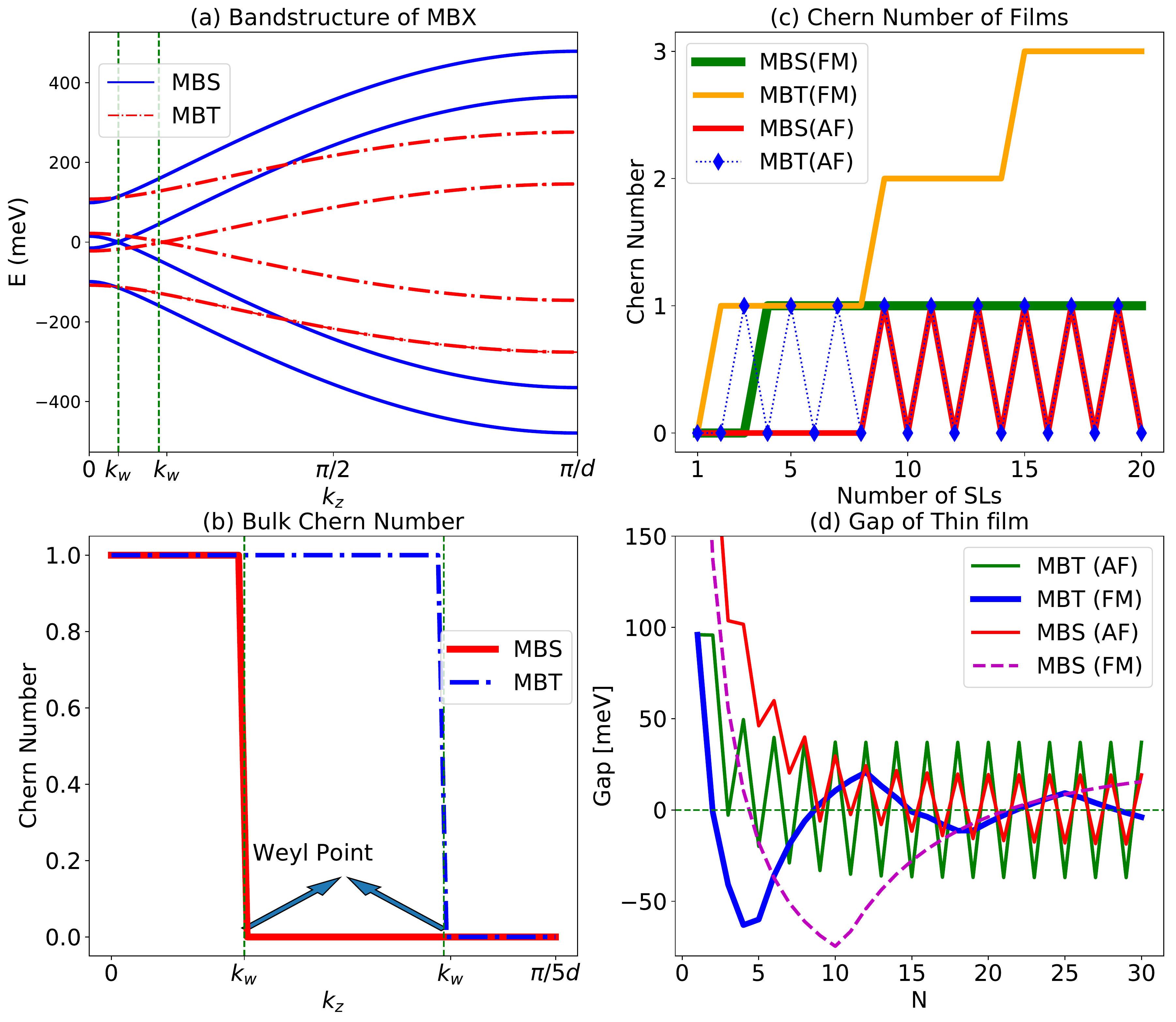}
\caption{(a) Bulk band dispersion of ferromagnetic MBX calculated from the 
Dirac cone model with the DFT-extracted parameters summarized in Table \ref{para_table}. 
Ferromagnetic MnBi$_2$X$_4$(X=Se,Te) is a Weyl semimetal, with the Weyl point located at 
$k_z = \pm k_w$ where $k_w \approx 3\pi/50d$ for MBS and $k_w \approx 3\pi/20d$ for MBT.
(b) Dirac-cone model $k_z$-dependent 2D Chern numbers for bulk MBX calculated 
from the model, showing jumps from 1 to 0 at the Weyl points.
(c) 2D Chern numbers of thin-film MBX calculated from the Dirac-cone model
which show that the Chern numbers of ferromagnetic MBX
jump to no-zero values beyond four septuple layers thickness for MBS and 2 
septuple layers for MBT.  The antiferromagnetic configurations 
have non-zero Chern numbers for odd-septuple-layer films
when the thickness is larger than 8 layers for MBS and 2 layers for MBT.
(d) Gap of thin-film MBX at the 2D-$\Gamma$ point as a function of the number of 
septuple layers.
}
\label{MBX} 
\end{figure}
\fi

The $\Gamma$ to $Z$ band dispersions calculated with these model parameters 
for the ferromagnetic configuration are illustrated in Fig.~\ref{MBX}(a).
Because the $\Gamma$-point gaps, around 30 meV for MBS and around 44 meV for MBT, are
inverted by the exchange splitting $m_{F}$, the spin-splitting closes at a 
finite value of $k_z$, as noted previously \cite{Li2019_theory,Chowdhury_2019},
generating a simple Weyl semimetal \cite{Wan2011_Weyl} with only two Weyl points. 
The model bands are in good agreement overall with the 
DFT results, although the small $\Gamma$-point gaps are even smaller in our DFT calculations,
which yield 23 meV for MBS and 12 meV for MBT. 
The role of longer-range hopping parameters $\Delta_D^n$, which are responsible for 
a velocity magnitude difference between the crossing bands at the Weyl point,
is addressed in SI Appendix, Fig. S2 by DFT calculations but dropped in the qualitative phase-diagram
discussions below since they are $\sim$ several meV and 
have little influence on the positions of Weyl points ($k_w$),
thin film AHEs, or 2D-$\Gamma$ point thin film gap (SI Appendix, Fig. S3) trends.
The bulk antiferromagnetic gaps calculated from this model are around 85 meV for MBS and 86 meV for MBT, 
compared to gaps estimated experimentally that vary from 50 meV to 200 meV\cite{nowka2017,Hu2020,Zeugner2019,Otrokov2019}.

In Fig.~\ref{MBX}(b) we plot 2D Chern numbers obtained by integrating the 
Berry curvature over $k_x$ and $k_y$ as a function of $k_z$, which are non-zero 
between the Weyl points at $k_z = \pm k_w$.
The Weyl momentum $k_w$ is non-zero for $m_{F} >  |\Delta_{D}|-|\Delta_{S}|$ and moves 
from $\Gamma$ to $Z$ with increasing $m_{F}$, reaching $\pi/d$ when $m_F=|\Delta_{D}|+|\Delta_{S}|$.  
The bulk Hall conductivity 
per layer is \cite{Burkov2018_Review} $k_w/(\pi/d) e^2/h$.  
Since the exchange interactions $m_F$ in MBX are larger than 
$|\Delta_{D}|-|\Delta_{S}|$, but quite small compared to $|\Delta_{S}|+|\Delta_{D}|$,
the Hall conductivity per septuple layer in bulk ferromagnetic MBX is small compared to the quantum
value $e^2/h$. 

In  Fig. \ref{MBX} (c) we plot the total Chern numbers of 
thin films with ferromagnetic and antiferromagnetic configurations 
as a function of the number of septuple layers in the film.  
For ferromagnetic configurations, the thin films have zero Chern number
for thicknesses up to a critical number of septuple layers, which is 4 for 
MBS and 2 for MBT.  As the thickness increases the Chern numbers increase 
indefinitely, increasing by one for every $\sim \pi/d/k_w $ added septuple 
layers so that the quantized Hall conductivity of very thick films 
approaches the bulk value when normalized per layer (See SI Appendix, Fig. S4).
As the Chern number 
increases the energy gaps tend to decrease, as illustrated in Fig. \ref{MBX} (d),
where we have assigned a negative sign to the gap for odd integer Chern numbers.
The oscillating gap size simply reflects the size-quantization  
of the bulk Weyl semimetal bands, as detailed in SI Appendix, Fig. S5. 
The behavior of antiferromagnetic configurations is quite distinct.
Here the Chern numbers are non-zero only for odd numbers of septuple layers 
and, in that case, only when a critical film thickness 
equal to 2 septuple layers for MBT and 8 septuple layers for MBS is exceeded.  
The antiferromagnetic configuration gaps are identical for even and odd numbers of 
layers in the thick film limit. (See SI Appendix, Fig. S5). 
These semi-analytic predictions of the Dirac cone model are consistent with 
recent experiments\cite{Deng2020,Ge2020,Liu2020} which have, in particular, 
shown that for MBT films the total Chern number jumps from 1 to 2 when the thickness reaches 
nine\cite{Ge2020} septuple layers. 

\section*{Superlattice Phase Diagrams}
The van der Waals character of MBX materials allows for property variation by changing the stacking sequence of magnetic septuple and non-magnetic quintuple 
layers\cite{Hagmann2017,Klimovskikh2019,Wu2019,Vidal2019,Hu2020}. 
Importantly, the magnetic field needed to convert between antiferromagnetic and 
ferromagnetic stacking arrangements can be reduced simply by inserting 
non-magnetic quintuple spacers between the magnetic septuple layers.
When magnetic and non-magnetic layers are simply alternated, for example,
it has been shown experimentally that the magnetic field needed to convert to 
a ferromagnetic configuration is reduced from $\sim 5$T to $\sim 0.22$T\cite{Hu2020}.

We consider the family of bulk crystals in which a template with 
$M$ MnBi$_2$X$_4$ septuple layers and $N$ Bi$_2$X$_3$ quintuple layers is repeated.  Several of these superlattices have 
already been realized experimentally including (M,N) = (1,1) (MnBi$_4$Se$_7$) \cite{Klimovskikh2019,Wu2019,Vidal2019,Hu2020},(M,N) = (1,2) (MnBi$_6$Se$_{10}$) \cite{Klimovskikh2019,Wu2019}, and
(M,N) = (1,3-6) ( MnBi$_8$Se$_{13}$, MnBi$_10$Se$_{16}$, and MnBi$_12$Se$_{19}$) \cite{Klimovskikh2019}.
In Fig. \ref{phase_stacking} we summarize the energy gaps 
and topological phases of ferromagnetic configuration
(MnBi$_2$X$_4$)$_M$/(Bi$_2$X$_3$)$_N$ superlattices.  
We find that the energy gap 
is mainly determined by the ratio of magnetic to non-magnetic layers $M/N$, and is less dependent on the order of the layers.  For example, 
when we set $\Delta_S^{MTI} = \Delta_S^{TI}$, we find that 
$M/N$ larger than around 3 leads to bulk 
Weyl semimetals.  Decreasing $\Delta_S^{MTI}/\Delta_S^{TI}$ favors 
Weyl semimetal phases with a decrease of around $\pm 10\%$
reducing the critical $M/N$ ratio to 2.  We therefore expect that stronger hybridization between surface states 
within the non-magnetic topological insulator layers favors Weyl semimetals.
Among (M,N) = (1,1) superlattices, candidates include MnBi$_2$Te$_4$ combined with 
Bi$_2$Se$_3$, Bi$_2$Sb$_3$, or Sb$_2$Te$_3$, which are estimated to have stronger 
same layer hybridization than Bi$_2$Te$_3$ according to the DFT calculations summarized in SI Appendix, Fig. S8.

\ifpdf
\begin{figure}[htp]
\includegraphics[width=0.9\linewidth]{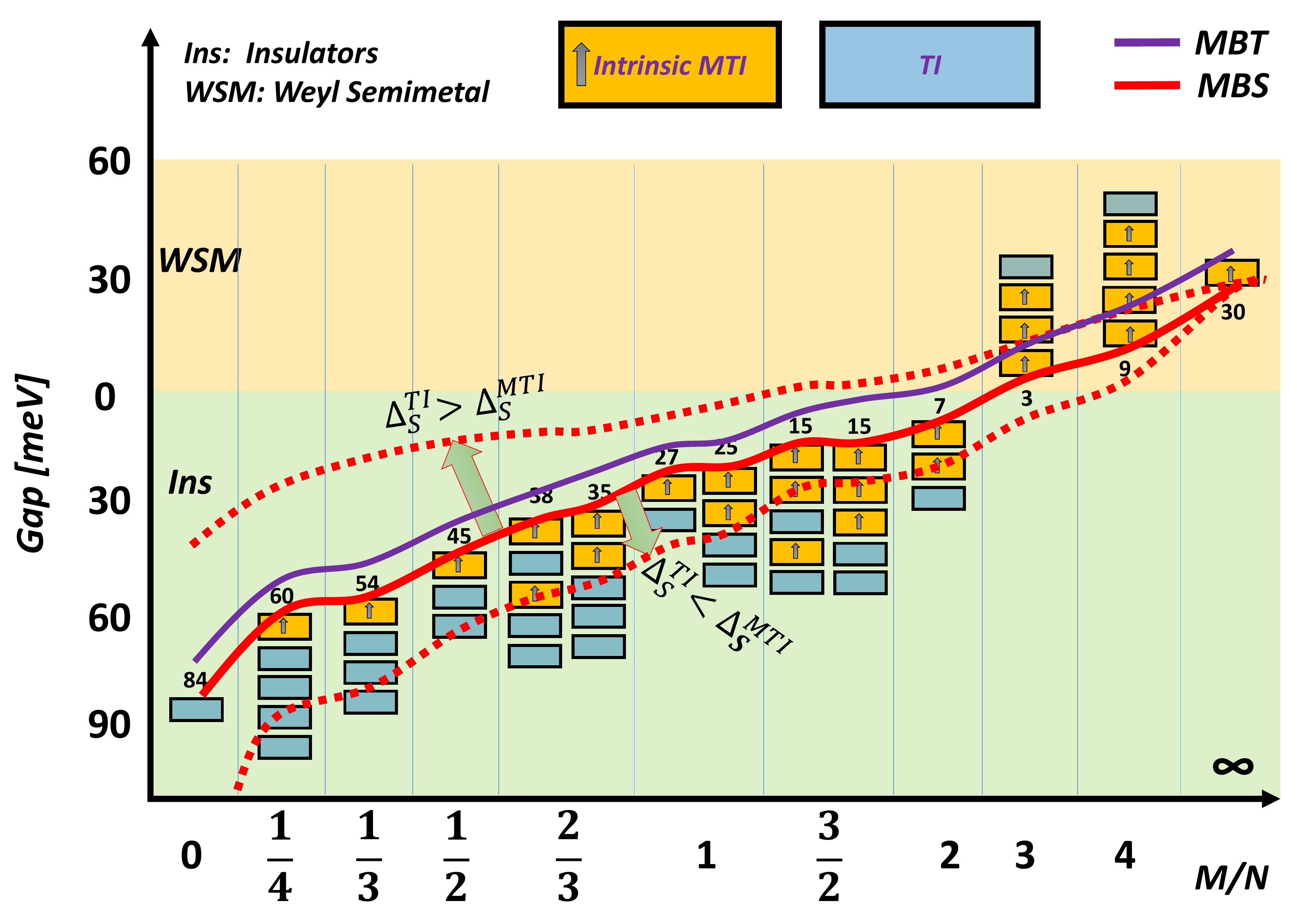}
\caption{\label{phase_stacking} 
Topological phase diagram of ferromagnetic-configuration (MnBi$_2$X$_4$)$_M$(Bi$_2$X$_3$)$_N$ superlattices {\it vs.}
$M/N$.  This figure was constructed by using the Dirac cone model 
to calculate energy bands and 
2D Chern numbers as a function of $k_z$ for all indicated superlattices.  
Weyl semimetal
superlattices are placed in the orange region of the phase diagram
and insulators in the green region.  MBS energy gaps (in meV) at 
the $\Gamma$ point
are listed at the top of the corresponding superlattice icons for insulating superlattices and the bottom for the
Weyl semimetal superlattices in which the gap has been inverted.  
The  solid red (purple) line plots $\Gamma$ point gaps for MBS (MBT)
calculated using $\Delta_S^{MTI} = \Delta_S^{TI}$, while the dashed red 
lines illustrate the cases in which $\Delta_S^{TI}$ deviates from $\Delta_S^{MTI} $ by $\pm$ 10 \%.
}
\end{figure}
\fi

The properties of (MnBi$_2$X$_4$)$_M$(Bi$_2$X$_3$)$_N$ superlattices and thin films
can be varied in a variety of different ways, for example by varying the 
pnictide fraction, applying pressure, or changing temperature.  
We study the efficacy of these tuning knobs by constructing phase diagrams {\it vs.} $J_S$ and $\Delta_{D}$,
as illustrated in Fig. \ref{phase_mag}, 
for several different stacking sequences keeping the ratio of 
$J_S$ to $J_{D}$ and $\Delta_{S}$ fixed.  To limit the number of parameters 
we used the same values for $\Delta_{S}$ and $\Delta_{D}$ in 
magnetic and non-magnetic layers.
The motivation for illustrating the dependence of phase 
on this particular subset of model parameters is that i) we expect
both exchange interactions to decline with temperature as the alignment of the 
local moment spins decreases with increasing temperature, and ii) that we expect 
the van der Waals gap to narrow as pressure is applied, increasing $\Delta_{D}$.
The dependence of $\Delta_{D}$ on pressure is addressed explicitly in SI Appendix, Fig. S7. 
For M = 1 and N = 0 (that is for bulk MnBi$_2$X$_4$), 
the phase diagram shown in Fig. \ref{phase_mag} (a) maps to that of the  
MTI/normal insulator superlattice model studied by Burkov and Balents \cite{Burkov_Balents_2011}. 
The cases of M/N = 0.5, 1 and 2 are shown in Fig. \ref{phase_mag} (b)-(d),
where the positions of MBS and MBT systems in the phase diagram are 
marked by red and blue dots. Although antiferromagnetic MnSb$_2$Te$_4$ is a trivial insulator, unlike MBT, according to DFT calculations\cite{Chen_2019_MST} and in agreement with our results, ferromagnetic MnSb$_2$Te$_4$ is a Weyl semimetal accroding to DFT calculations shown in SI Appendix Fig. S2. The estimated model parameters are $m_F = J_S+J_D \approx 45 meV$, $\Delta_S \approx 124 meV$ and $\Delta_D \approx -166 meV$ for MnSb$_2$Te$_4$ and thus it should lie between that of MBT and and MBS in the phase diagram of Fig. \ref{phase_mag}.

In all cases the transition
between normal insulator and Weyl semimetal states occurs at weakest 
exchange coupling near $\Delta_D = \Delta_S $, 
which marks the boundary between normal insulator and 
topological insulator states in the non-magnetic $J_{S/D}=0$ limit.
For the limit $\Delta_D = 0$, which corresponds to isolated layers, 
a phase transition from a trivial insulator to a Chern insulator state occurs
when the exchange interaction exceeds $\Delta_{S}$; the 
Weyl states emerge at smaller $J_{S}$ for interior magnetic layers  
because both exchange interactions\ $J_{S}$ and $J_{D}$ contribute.  
For example for $M=1,N=0$, illustrated in Fig. \ref{phase_mag} (a), the 
band energy at the $\Gamma$ point is 
$E_{\Gamma} = \pm (J_S+J_D) \pm \Delta_S$, and 
the phase transition to the Chern insulator occurs when 
$J_S+J_D > \Delta_S$, that is at  $J_S=\Delta_S/(1+\delta)$ with $\delta \equiv J_D/J_S $. 
For the case of $M=1,N=1$ shown Fig. \ref{phase_mag} (b), the 
magnetic layers are isolated so that 
$ E_{\Gamma} = \pm J_S \pm \Delta_S $ and $\pm J_D \pm \Delta_S$, and there are therefore two phase transition points at $\Delta_{D}=0$ one at $J_S = \Delta_S $ and one at $ J_D = \Delta_S $.
When the latter phase transition occurs the total Chern number per period 
changes from $C=1$ to $C = 2$.  At finite $\Delta_{D}$ quantum Hall states with different 
Chern numbers per period are separated by Weyl semimetal states.  
Fig. \ref{phase_mag} (c) and (d) show the cases of $M=1,N=2$ and $M=2,N=1$
for which the eigenvalues in the $\Delta_D$ to 0 limit are discussed in 
SI Appendix and imply critical values of $J_S = \Delta_S $ for the $M=1,N=2$
case, and $J_S = \Delta_S/\sqrt{1+\delta}, \Delta_S/\delta $ 
for the case of $M=2,N=1$ case.
It shows that ferromagnetic Mn$_2$Bi$_6$Te$_{11}$ has large chance to be a Weyl semimetal from Fig. \ref{phase_mag} (d).

\ifpdf
\begin{figure}[htp]
\includegraphics[width=0.9\linewidth]{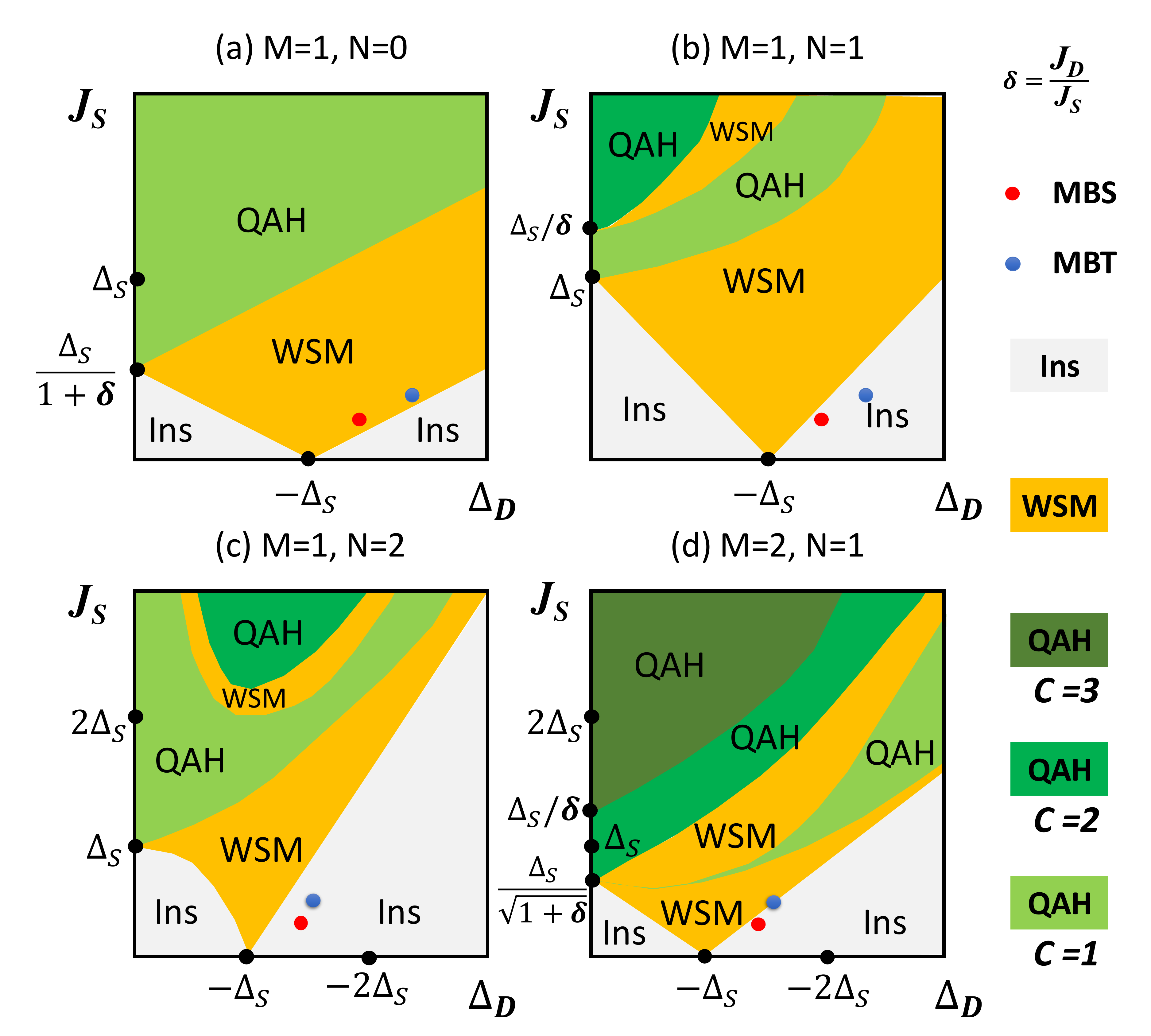}
\caption{\label{phase_mag} 
Phase diagram of ferromagnetic (MnBi$_2$X$_4$)$_M$ / (Bi$_2$X$_3$)$_N$ superlattices with exchange interactions $J_S(J_D)$ and Dirac cone coupling energies 
$\Delta_D$ in adjacent layers and $\Delta_S$ in the same layer, where (a)-(d) are for (M,N) = (1,0)(MnBi$_2$X$_4$), (M,N) = (1,1)(MnBi$_4$X$_7$), 
(M,N) = (1,2)( MnBi$_6$X$_{10}$), and (M,N) = (2,1)(Mn$_2$Bi$_6$X$_{11}$).
Increasing temperature moves points down in this phase diagram whereas applying pressure moves points to the right.  The zero-temperature MBS and MBT parameters are marked by red and blue dots. From (d) it shows that ferromagnetic Mn$_2$Bi$_6$Te$_{11}$ has large chance to be a Weyl semimetal.
}
\end{figure}
\fi

\section*{Discussion}
The coupled Dirac cone model developed here provides
an excellent qualitative description of MnSb$_x$Bi$_{2-x}$X$_4$ (MPX)
layered van der Waals materials in both thin film and bulk crystal limits.
The simplified Dirac cone model has the advantage that it can readily be used 
as a platform to address the influence of gating, disorder, and external magnetic fields,
all relevant perturbations that are not easily accounted for using 
\textit{ab initio} approaches.  It can also help explain trends across the 
material family, and simplify theories of systems with less translational symmetry
(films that are many layers thick or bulk crystals with complex stacking arrangements)
and theories of interaction effects.  We anticipate that in both ferromagnetic and 
antiferromagneric thin films, gating can be used to 
drive transitions between insulators with different total Chern numbers, including 
between quantum anomalous Hall insulators with non-zero total Chern numbers and 
ordinary insulators with total Chern numbers equal to zero.  In 
bulk crystals we anticipate the simplified model will enable quantitative 
descriptions of the chiral magneto-transport anomaly \cite{Son2013,Burkov2014}
of ferromagnetic states and the axion magneto-electric response of even layer 
antiferromagnetic states.

We have used the Dirac cone model here to explain the main trends in topological properties, namely that i) bulk antiferromagnetic MPX compounds are topological insulators,
ii) bulk ferromagnetic MPX compounds are Weyl semimetals, iii) thin-film antiferromagnets are 2D Chern insulators when the number of layers is odd and exceeds a critical value and 
ordinary insulators otherwise, and iv) thin film
ferromagnets that exceed a critical thickness are 2D Chern insulators whose Chern numbers grow and 
gaps decline with film thickness.  We have reached these conclusions based 
mainly on the properties of van der Walls heterojunctions with
MBT and MBS building blocks, but based on the phase diagrams in 
Fig.~\ref{phase_mag} they appear to be fairly robust.
We have used the model to verify that it is not possible to realize 
Weyl semimetals by a repeated stacking of odd layer number antiferromagnets
separated by non-magnetic spacers, even though they are 2D Chern insulators when isolated.  
To move more deeply into the topologically non-trivial part of
phase diagram, the best prospect seems to be in engineering constituent layers 
so as to increase $\Delta_S$ or decrease $|\Delta_D|$
so that they are nearly equal in magnitude, for example by tuning the pnictide fraction.

Since the MPX compounds are antiferromagnetic in their ground states, it is 
necessary to apply an external magnetic field to align the moments and reach the
ferromagnetic configuration.  Because the antiferromagntic exchange 
interaction across the van der Waals gap is weak, 
the magnetic field needed to align the magnetic moments is relatively small 
($\sim 5$ T).  The field strength necessary to align moments in layers separated 
by a single non-magnetic spacer layer is much smaller ($\sim 0.2$ T).
It follows that there is a broad field range in which the ground state 
configuration will consist of units with no spacers within which the moments are 
arranged antiferromagnetically, with the net spin of each unit aligned with the field.  
Still more complex magnetic configurations should be reachable in even weaker fields 
with suitably chosen stacking arrangments that include double non-magnetic spacers.

It is interesting to consider the implications of the Dirac cone model 
for the interpretation of photoemission experiments, which are confusing at present.
Some ARPES experiments\cite{Li2019,Hao2019,Chen2019,Swatek2020}, generally those 
using small photon energies near $\sim 7$ eV have not observed the surface-state gaps
predicted for thick antiferromagnetic films 
by both DFT and by the coupled Dirac-cone model.  
Gapped Dirac cones were observed in higher 
photon energy (usually larger than 20 eV) ARPES experiments \cite{Zeugner2019,Otrokov2019,Vidal2019b,Lee2019}, which are likely less surface sensitive.  
One possible explanation is that the lower energy ARPES experiments are more 
sensitive to the topologically protected
edge states expected on the antiferromagnetic state surface at step edges where the number of layers present 
in the system changes.  As illustrated in SI Appendix, Fig. S6, adding a layer to a MTI antiferromagnet either changes 
the Hall conductivity from zero to $\pm e^2/h$ or from $\pm e^2/h$ to zero.  In either case topologically protected states are required by the quantized change in Hall conductivity.

\matmethods{
\subsection*{DFT Calculations}
The DFT calculations summarized below were implemented in the Vienna Ab initio simulation package(VASP)~\cite{vasp}, using 
semi-local PBE-GGA~\cite{pbe}.  When Mn atoms were present 
a Hubbard on-site electron-electron interaction\cite{LDAU_Himmetoglu_2013}(LDA+U)  
with U = 5.25 eV was included for its 3d electrons. 
The cutoff energy for the plane-wave basis was set to be 600 eV, 
the global break condition for the electronic self-consistency loop was set to $10^{-7}$ eV and  
the mesh size for creating the k-point grid was set to $9\times9\times3$ for bulk calculations and ($9\times9\times1$ for thin films).  
For MBS and MBT the lattice constants were obtained 
by relaxing the unit cell until
the forces for each atom were smaller than $ 1 \times 10^{-3}$ eV/ $\angstrom$.
}

\showmatmethods{} 

\acknow{This work Research was sponsored by the Army Research Office under Grant Number W911NF-16-1-0472.
, and by the Welch Foundation under grant TBF1473. We acknowledge generous 
computer time allocations from the Texas Advanced Computing Center.}

\showacknow{} 

\bibliography{mbx}

\pagebreak

\onecolumn

\setcounter{equation}{0}
\setcounter{figure}{0}
\setcounter{table}{0}
\setcounter{page}{1}
\renewcommand{\theequation}{S\arabic{equation}}
\renewcommand{\thefigure}{S\arabic{figure}}
\renewcommand{\thetable}{S\arabic{table}}
\renewcommand{\bibnumfmt}[1]{[S#1]}
\renewcommand{\citenumfont}[1]{S#1}

\begin{center}
\title{ \begin{huge} 
\textbf{Supplemental Materials for \\
"Magnetized Topological Insulator Multilayers"} 
\end{huge} }
\end{center}

\subsection*{Model parameters from bulk DFT calculations}
We estimate the model Dirac velocity ($v_{_D}$) parameters 
by examining the dependence of DFT band energies on 
$\mathbf{k}_{\perp}$, which are illustrated in Fig. \ref{band_kperb}.
For each $k_z$, the model energy bands for ferromagnetic configurations are given by
\begin{equation}
E(\mathbf{k}_{\perp},k_z) = \pm \sqrt{m^2 + (\hbar v_D \mathbf{k}_{\perp})^2},
\end{equation}
where $m = E(\mathbf{k}_{\perp}=0,k_z) $ is the mass at $k_z$.
The DFT results suggest that the approximation of 
linear Dirac dispersion is reasonable within several hundreds of meVs of the Fermi energy.
The linear Dirac dispersion is hidden in the non-magnetic case by the 
large gaps already present at $\mathbf{k}_{\perp}=0$.    
The resulting Dirac velocity parameters are summarized in Table \ref{para_table} and 
lie in the range of a few times $10^5$m/s, which is typical of Dirac materials.

\ifpdf
\begin{figure}[htp]
\centering
\includegraphics[width=0.9 \textwidth]{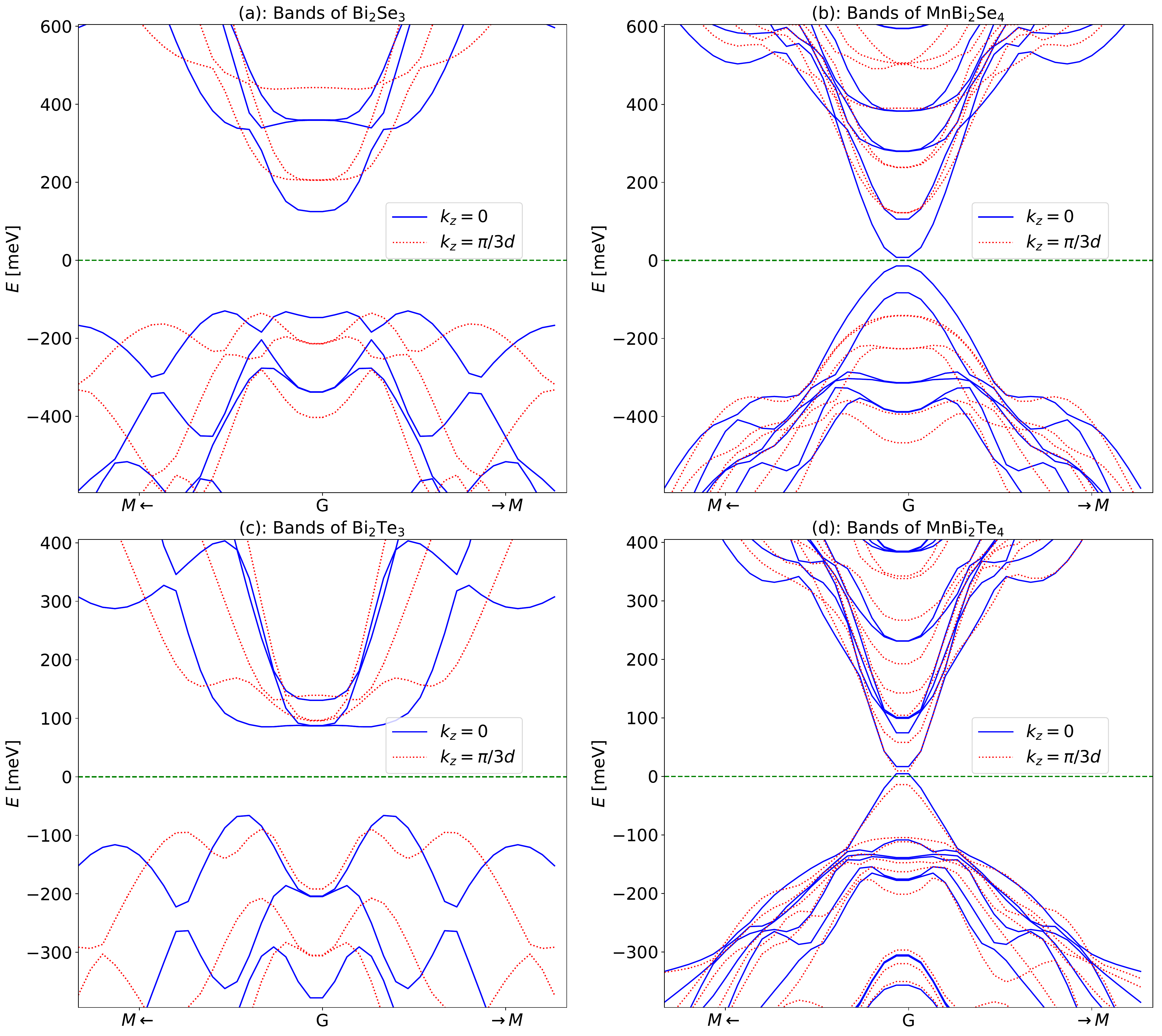}
\caption{Ferromagnetic configuration DFT bandstructure \textit{vs.} $\mathbf{k}_{\perp}$.
The solid blue lines were calculated at $k_z = 0$ and the red dashed lines at $k_z = \pi/3d$. 
(a): Bands of Bi$_2$Se$_3$;
(b): Bands of MnBi$_2$Se$_4$;
(c): Bands of Bi$_2$Te$_3$;
(d): Bands of MnBi$_2$Te$_4$. }\label{band_kperb}
\end{figure}
\fi

\begin{table}\centering
\caption{Fermi velocities and lattice constants from DFT calculations.
The range of velocities reflects a combination of anisotropy and 
uncertainly in the fitting procedure.}\label{para_table}
\begin{tabular}{llc}
Materials & Fermi velocity (m/s) & lattice constant($\angstrom$)\\
\midrule
Bi$ _2$ Se$ _3 $ & 5.2 $\times$ 10$^5$ - 7.2 $\times$ 10$^5$ & 4.138\\
MnBi$ _2$ Se$ _4 $ & 3.7 $\times$ 10$^5$ - 5.6 $\times$ 10$^5$ & 4.078 \\
Bi$ _2$ Te$ _3 $ & 3.6 $\times$ 10$^5$ - 5.1$\times$ 10$^5$ & 4.383 \\
MnBi$ _2$ Te$ _4 $ & 3.9 $\times$ 10$^5$ - 5.5 $\times$ 10$^5$ & 4.407 \\
\bottomrule
\end{tabular}
\end{table}

DFT band structures along the $\Gamma$ to $Z$ lines of the ferromagnetic configurations of 
MnBi$_2$Te$_4$, MnBi$_2$Se$_4$, MnSb$_2$Te$_4$ and MnBi$_2$Se$_2$Te$_2$ are illustrated in
Fig. \ref{band_kz}.  In the last case the two middle Te atoms were replaced by Se atoms.
These DFT band structure bands are folded three times since the structure repeats only after 
three septuple units when the microscopic stacking arrangement is taken into account.
Note that MnBi$_2$Se$_2$Te$_2$ is a normal insulator, not a Weyl semimetal.

\ifpdf
\begin{figure}
\centering
\includegraphics[width=0.9 \textwidth]{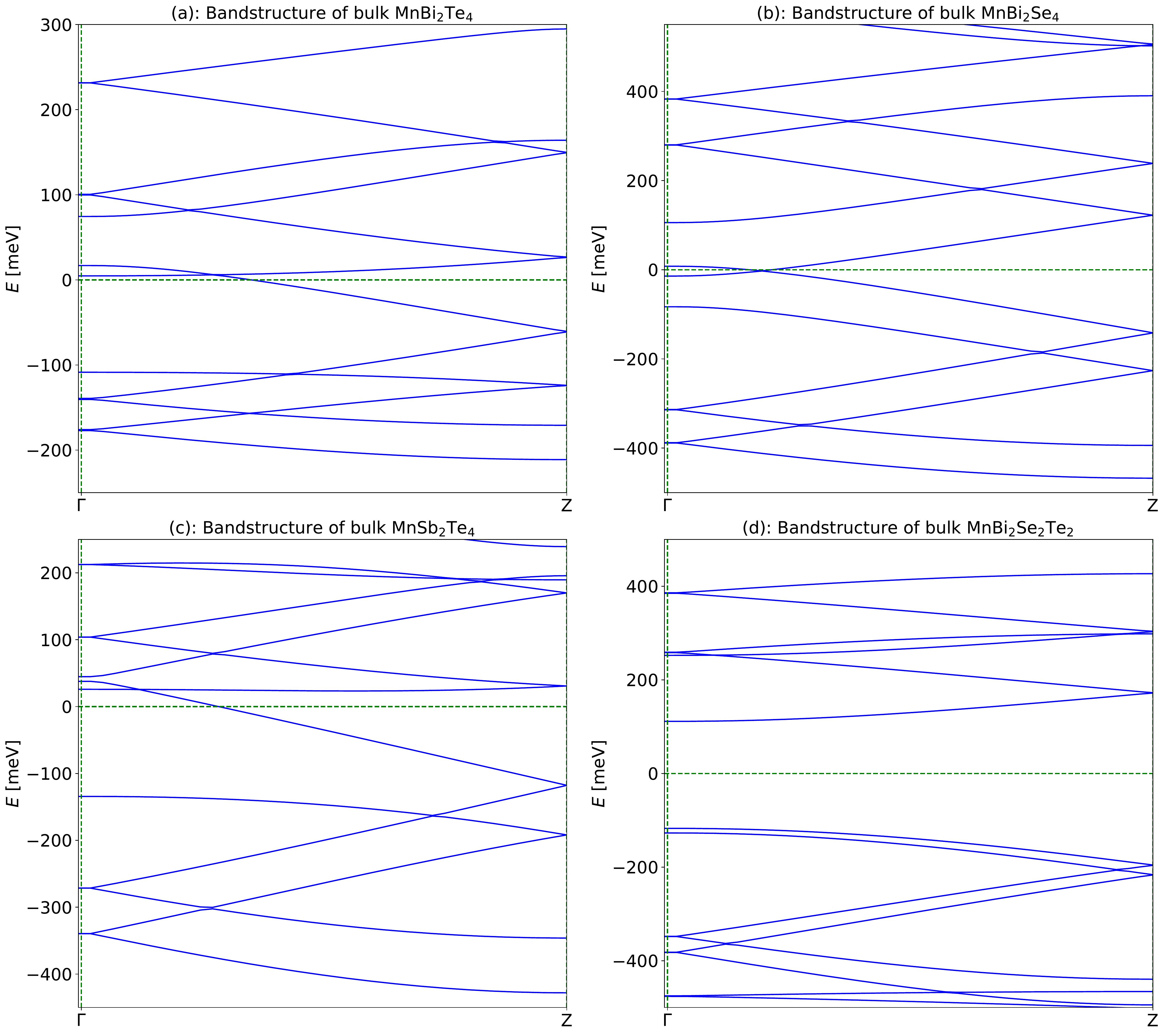}
\caption{Ferromagnetic configuration DFT bandstructures \textit{vs.} $\mathbf{k}_{z}$ from $\Gamma$ to Z
at $\mathbf{k}_{\perp}=0$.  Since the microscopic structure repeats only after 3 septuple units, the and Z point lies at 
(0,0,$\pi/3d$) where $d$ is the septuple layer separation. 
(a): Bands of ferromagnetic MnBi$_2$Te$_4$;
(b): Bands of ferromagnetic MnBi$_2$Se$_4$;
(c): Bands of ferromagnetic MnSb$_2$Te$_4$(MST), in which the Bi atoms are replaced by Sb atoms, the estimation of model parameters that fit the MST DFT bandstructure are $m_F = J_S+J_D \approx 45 meV$, $\Delta_S \approx 124 meV$ and $\Delta_D \approx -166 meV$;
(d): Bands of ferromagnetic MnBi$_2$Se$_2$ Te$_2$, in which the inner two Te atoms in each septile unit
are replaced by Se atoms.}\label{band_kz}
\end{figure}
\fi

In Fig. \ref{gap_film_para} we compare DFT thin film gaps at the 
2D $\Gamma$ point {\it{vs.}} septuple layer number with 
model gaps {\it{vs.}} for both ferromagnetic and antiferromagnetic configurations.
The model calculations were performed using the simplest version of the model with only the  
four largest parameters ($J_S$, $J_D$, $\Delta_S$, $\Delta_D$) which were extracted from
bulk band energies at $k_z=0$ and $k_z=\pi/d$.  
We have assigned a negative sign to the band gap when the thin film has an odd Chern number.
Although for the gaps calculated with the model are less accurate for antiferromagnetic configurations,
they do capture the trends reasonably well.  In all cases, the quality of the fit to DFT can be improved by adding more model paramaeters, such as $\Delta_D^1$, $\Delta_D^2$, $\Delta_D^3$ defined in the main text. 

\ifpdf
\begin{figure}
\centering
\includegraphics[width=0.9 \textwidth]{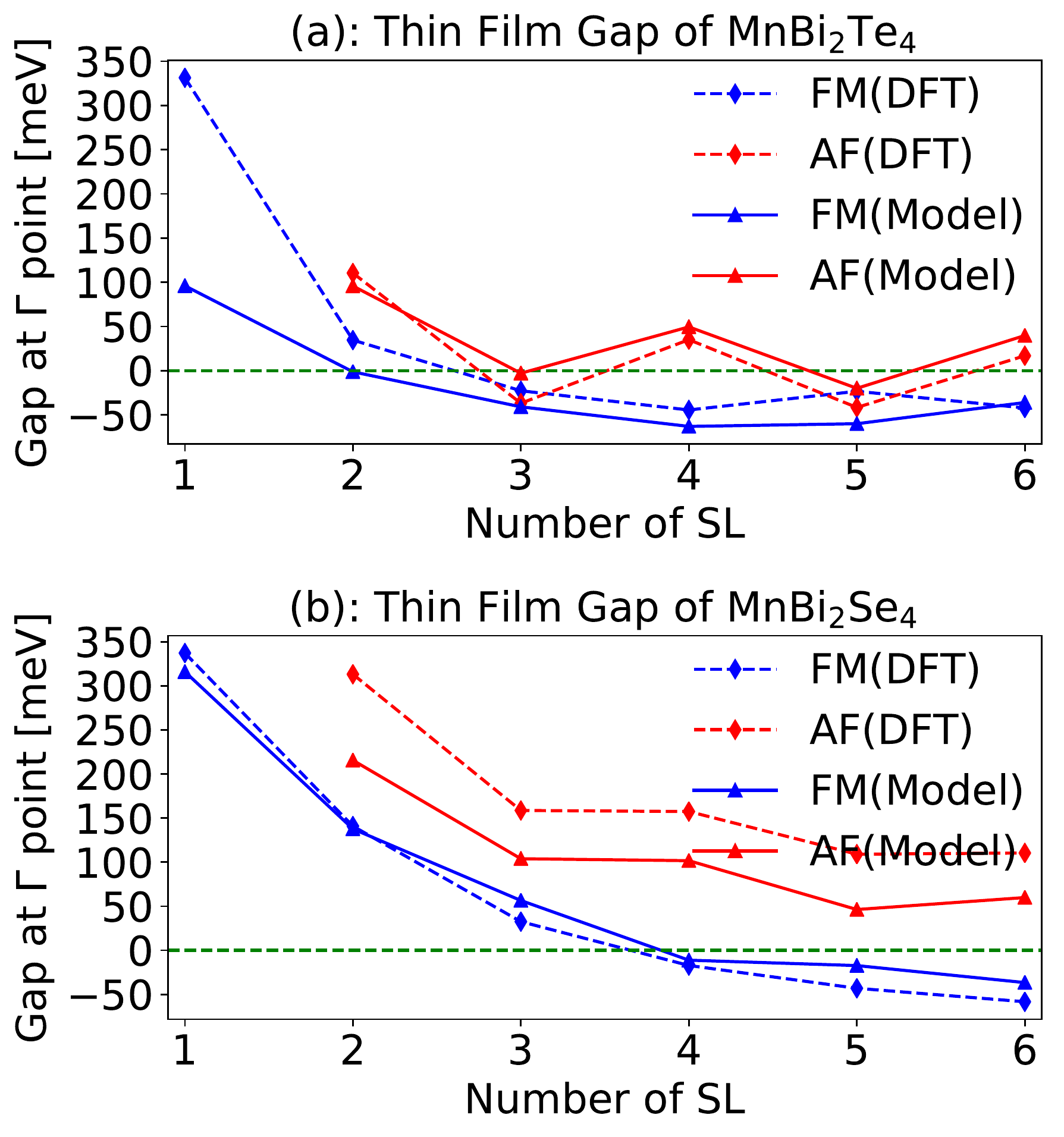}
\caption{Comparison between model and DFT thin film 2D $\Gamma$ point gaps.
(a) is for MnBi$_2$Te$_4$ (MBT) and
(b) is for MnBi$_2$Se$_4$ (MBS).
The signs of gap are assigned based on the 
Chern numbers C, with negative signs applied to odd Chern numbers.
}\label{gap_film_para}
\end{figure}
\fi

\subsection*{Thin film Chern numbers}
Chern numbers were calculated by integrating the Berry curvature 
\begin{equation}
    \Omega_n(\mathbf{k}) = -2 \sum_{n' \ne n} Im\Big[\frac{<n|\partial_{k_x} H_{\mathbf{k}}|n'><n'|\partial_{k_y} H_{\mathbf{k}}|n>}{(E_n - E_{n'})^2}\Big]
\end{equation}
over $k_x$ and $k_y$.  The Chern number of the n$^{th}$ subband is 
\begin{equation}
    C_n = \frac{1}{2\pi}\int d \mathbf{k} \Omega_n(\mathbf{k})
\end{equation}
and the total Chern number below Fermi level $E_F$ is
\begin{equation}
    C = \sum_{E_n < E_F} C_n =  -\frac{1}{\pi} \sum_{E_n < E_F,E_{n'}>E_F} \int d \mathbf{k}  Im\Big[\frac{<n|\partial_{k_x} H_{\mathbf{k}}|n'><n'|\partial_{k_y} H_{\mathbf{k}}|n>}{(E_n - E_{n'})^2}\Big] .
\end{equation}

In Fig.  \ref{chern_weyl} we plot thin film Chern numbers normalized per septuple layer.
In this calculation we fix the Dirac cone hybridization parameters $\Delta_S$ and $ \Delta_D$ to 
be 84 meV and -127 meV, corresponding to MBT. As shown in the main text, MBX is a 
Weyl semimetal when $|\Delta_D| - |\Delta_S| < J_S + J_D < |\Delta_D| + |\Delta_S|$. 
In Fig. \ref{chern_weyl} we have varied $m_{F}=J_S + J_D$ 
from 40 meV to 220 meV.  Over this range of exchange interactions strengths 
the bulk Weyl point varies $k_z = 0$ to $k_z = \pi/d$ so that the bulk Hall conductivity 
normalized per layer,
\begin{equation}
   \frac{\sigma^{3D}_{xy}}{d} = \frac{e^2}{h} \; \frac{k_w d}{\pi}
\end{equation}
varies from $0$ to $e^2/h$.  The corresponding thin film hall conductivities per layer 
are plotted in Fig.  \ref{chern_weyl} for film thicknesses ranging from 1 to 12 septuple layers. 
Fig.  \ref{chern_weyl} shows that the thin film Chern numbers per septuple layer
approaches the bulk value as the film thickness increases.

\ifpdf
\begin{figure}
\centering
\includegraphics[width=0.9 \textwidth]{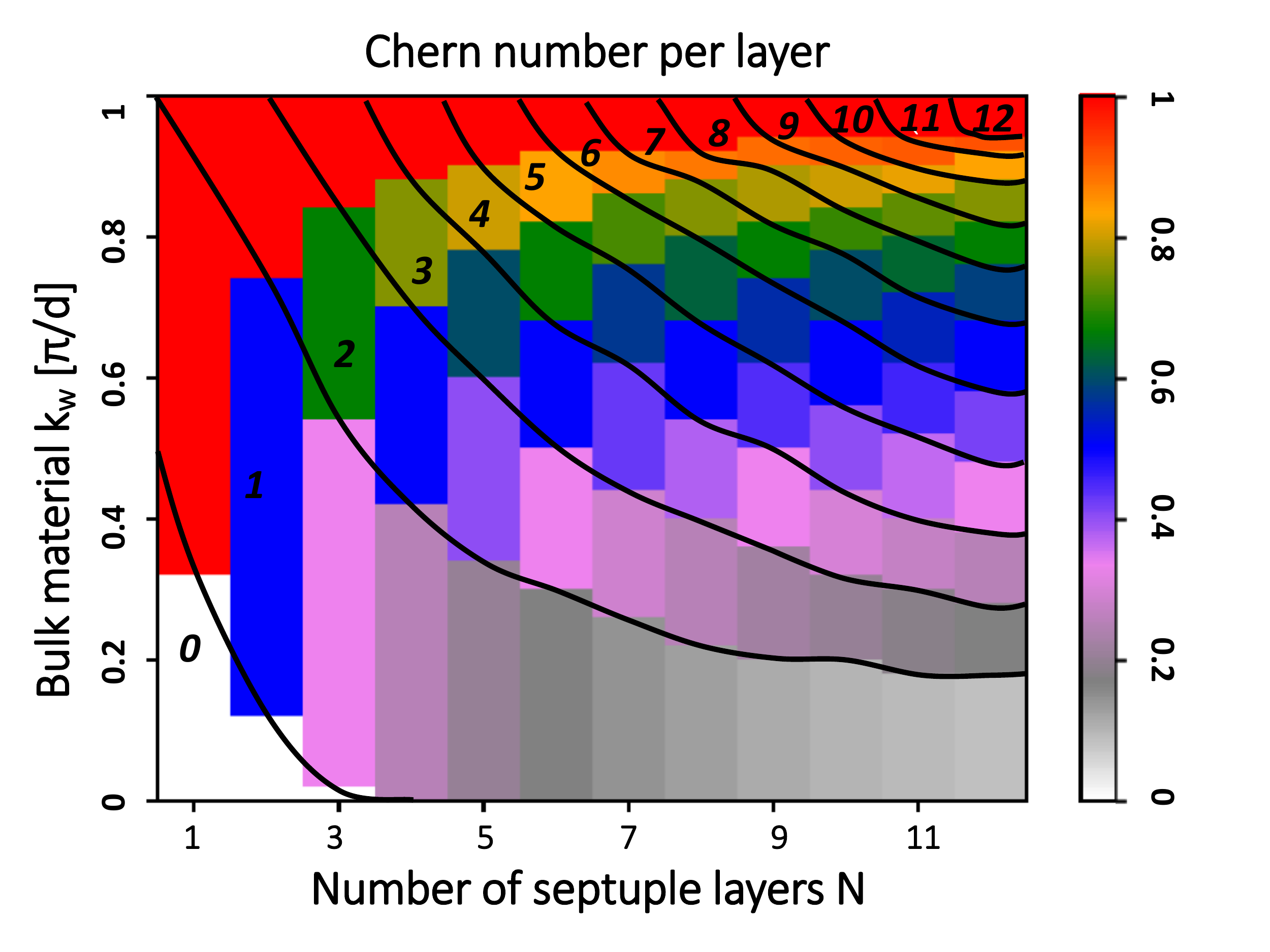}
\caption{Chern numbers per septuple layer (color scale)
for ferromagnetic thin films \textit{vs.} 
number of layers from 1 to 12 with the strength of the ferromagnetic exchange 
$m_{F}=J_{S}+J_{D}$ adjusted so as to drive the bulk Weyl momentum $k_w$ from 
the $\Gamma$-point to the zone boundary.  Other model parameters 
retain their MnBi$_2$Te$_4$ values. The Chern number per layer is obtained
by dividing the total Chern number (indicated by black numbers, with 
black lines separating different regimes) by the number of layers.
The Chern number per layer approaches its bulk value $k_w d/\pi$ as the film thickness increases.}\label{chern_weyl}
\end{figure}
\fi

\subsection*{Thin-film band energy trends}
The evolution of thin film low-energy bands of MBX at the 2D $\Gamma$ point with film thickness
is illustrated in Fig.\ref{band_energy_layers} for both ferromagnetic and antiferromagnetic cases 
and for X=Se and X=Te.  For the ferromagnetic configuration, the bulk systems are 
Weyl semimetals. Size-quantization of the semimetal bands leads to a set of level 
crossings between opposite spin states at which Chern numbers change as discussed in the bulk.
For  MnBi$_2$Te$_4$ these crossings occur near N = 2, 9, 15, 22, 28 \textit{etc.},
whereas for MnBi$_2$Se$_4$ which has a smaller bulk Weyl momentum 
the crossings occur near N = 4, 22, \textit{etc.}. 
For antiferromagnetic states, a level crossing occurs near N=3 for 
MnBi$_2$Te$_4$ and near N=9 for MBS.  Beyond this level crossing the 
total Chern number is one for odd numbers of layers and zero for even numbers of 
layers. In the antiferromagnetic case the lowest two energy states in thick films are polarized 
 at the surface. The crossover from normal to Chern (odd N) and axion insulator (even N)
 states occurs close to where the spin-splitting at the surface becomes larger 
 than the surface-state splitting in the non-magnetic state. 

In Fig. \ref{gap_delta} (a)-(d) we illustrate how the low-energy spectrum at 
$ \Gamma $ changes upon addition of a single layer by plotting the lowest energy 
states of both ferromagnetic and antiferromagnetic thin films as the coupling $\Delta_D$ to 
the added septuple layer is increased from zero, and as the coupling $\Delta_S$ between the top and bottom
surfaces of the added layer is increased from zero.
In Fig. \ref{gap_delta} (b), we illustrate the variation of 
low energy eigenvalues as the coupling $\Delta_{D}$ between an antiferromagnetic
N-1 layer thin film and a single septuple layer is increased from zero to its experimental value. 
For N larger than the critical thickness, there is always a gap closing as $\Delta_{D}$ is 
increased, consistent with the difference in Chern number between
even and odd layer number films.  Once the layer has been added, however, the gap returns to the 
same value it had prior to layer addition if the film is sufficinetly thick;  
the electronic structure at the surface of a 
thick film is not sensitive to whether the number of layers is even or odd; that is to say that the 
surface electronic structures of Chern insulator and axion insulator states are identical.  
The difference between these states is manifested only on the side walls of 
finite cross-sectional area films.  In the ferromagnetic case, level crossings occur only
as layers are added only when the $N$ and $N-1$ septuple layer films have different Chern numbers.  

\ifpdf
\begin{figure}
\centering
\includegraphics[width=0.9 \textwidth]{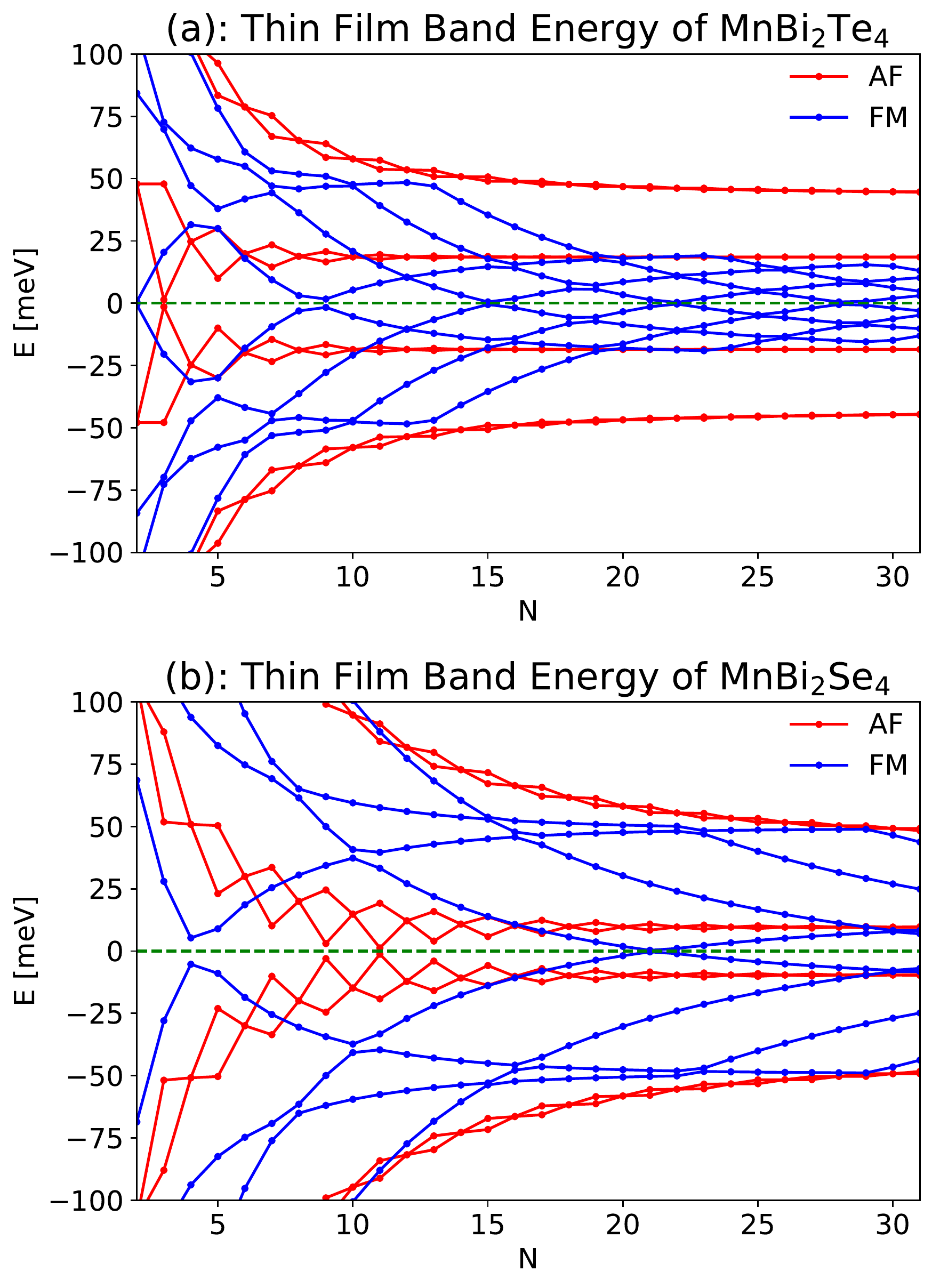}
\caption{The eight 2D $\Gamma$-point band energies closest to 
the Fermi level of neutral thin films of MnBi$_2$X$_4$ (MBX).  
The red solid lines are for the case of antiferromagnetic configurations 
and the blue solid lines for the case of ferromagnaetic 
configurations.
(a): Band energies vs number of layers for MnBi$_2$Te$_4$ (MBT);
(b): Band energies vs number of layers for MnBi$_2$Se$_4$ (MBS).
}\label{band_energy_layers}
\end{figure}
\fi

\ifpdf
\begin{figure}
\centering
\includegraphics[width=0.9 \textwidth]{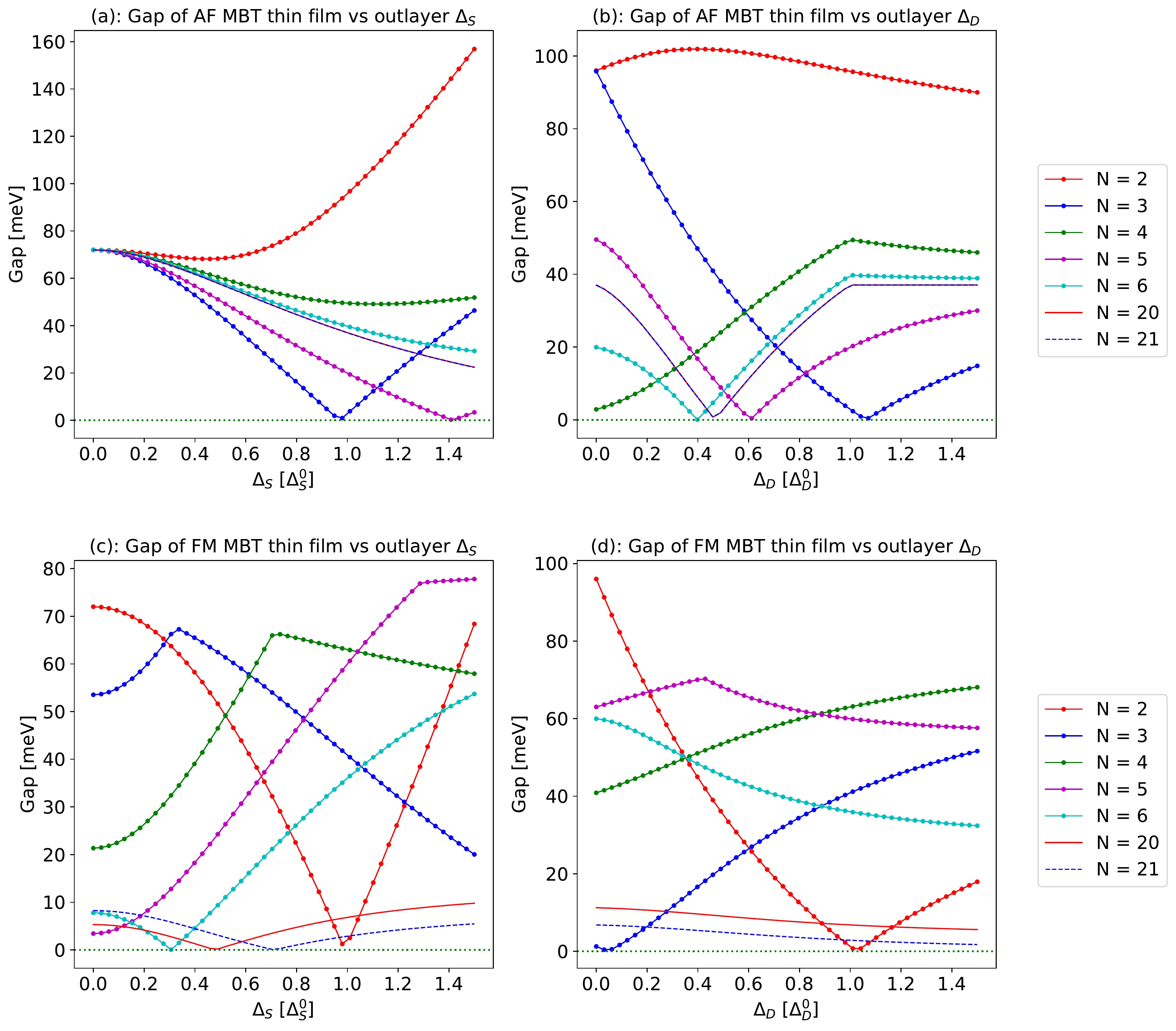}
\caption{Gap \textit{vs.} coupling to added Dirac cones.
(a): 2D Gap at the $ \Gamma $ point for antiferromagnetic MnBi$_2$Te$_4$ (MBT)
thin film \textit{vs.} the coupling $\Delta_{S}$ between Dirac cones on top and bottom of
the top layer.
(b): 2D Gap at the $ \Gamma $ point for antiferromagnetic MnBi$_2$Te$_4$ (MBT)
thin film \textit{vs.} the coupling $\Delta_{D}$ between an N-1 septuple layer film 
and the top septuple layer.
(c): 2D Gap at the $ \Gamma $ point for ferromagnetic MnBi$_2$Te$_4$ (MBT)
thin film \textit{vs.} the coupling $\Delta_{S}$ between Dirac cones on top and bottom of
the top layer.
(d): 2D Gap at the $ \Gamma $ point for ferromagnetic MnBi$_2$Te$_4$ (MBT)
thin film \textit{vs.} the coupling $\Delta_{D}$ between an N-1 septuple layer film 
and the top septuple layer.
}\label{gap_delta}
\end{figure}
\fi

\subsection*{Phase diagram vs stacking}
The topological character of bulk superlattice states (MBX)$_M$(BX)$_N$ 
can usually be determined by examining the band energy ordering at the 3D $\Gamma$-point.
In some limits of the model these are easily calculated analytically.  
In the case of $\Delta_D = 0$,
for M = 1, N = 0,
the energies at $\Gamma$ point are:
\begin{equation}
E = \pm (J_S+J_D) \pm \Delta_S
\end{equation}
For M = 1, N = 2,
the energies at $\Gamma$ point are:
\begin{equation}
\begin{split}
&E = \pm J_S \pm \Delta_S; \\
&E = \frac{1}{2} \Big( \pm J_D \pm \sqrt{J_D^2+4\Delta_S^2}\Big) ~~ (2 ~fold ~ degenerate)
\end{split}
\end{equation}
For M = 1, N = 1,
the energies at $\Gamma$ point are:
\begin{equation}
\begin{split}
&E = \pm J_S \pm \Delta_S; \\
&E = \pm J_D \pm \Delta_S
\end{split}
\end{equation}
For M = 2, N = 1,
the energies at $\Gamma$ point are:
\begin{equation}
\begin{split}
&E = \pm J_D \pm \Delta_S; \\
&E = \frac{1}{2} \Big( \pm(2J_S+J_D) \pm \sqrt{J_D^2+4\Delta_S^2}\Big) ~~ (2 ~fold ~ degenerate)
\end{split}
\end{equation}
For M = 3, N = 1,
the energies at $\Gamma$ point are:
\begin{equation}
\begin{split}
&E = \pm J_D \pm \Delta_S; \\
&E = \pm (J_S+J_D) \pm \Delta_S; \\
&E = \frac{1}{2} \Big( \pm(2J_S+J_D) \pm \sqrt{J_D^2+4\Delta_S^2}\Big) ~~ (2 ~fold ~ degenerate)
\end{split}
\end{equation}
For M = 4, N = 1,
the energies at $\Gamma$ point are:
\begin{equation}
\begin{split}
&E = \pm J_D \pm \Delta_S; \\
&E = \pm (J_S+J_D) \pm \Delta_S; ~~ (2 ~fold ~ degenerate) \\
&E = \frac{1}{2} \Big( \pm(2J_S+J_D) \pm \sqrt{J_D^2+4\Delta_S^2}\Big) ~~ (2 ~fold ~ degenerate)
\end{split}
\end{equation}
For M = 5, N = 1,
the energies at $\Gamma$ point are:
\begin{equation}
\begin{split}
&E = \pm J_D \pm \Delta_S; \\
&E = \pm (J_S+J_D) \pm \Delta_S; ~~ (3 ~fold ~ degenerate) \\
&E = \frac{1}{2} \Big( \pm(2J_S+J_D) \pm \sqrt{J_D^2+4\Delta_S^2}\Big) ~~ (2 ~fold ~ degenerate)
\end{split}
\end{equation}
For M = M, N = 1,
the energies at $\Gamma$ point are:
\begin{equation}
\begin{split}
&E = \pm J_D \pm \Delta_S; \\
&E = \pm (J_S+J_D) \pm \Delta_S; ~~ (M-2 ~fold ~ degenerate) \\
&E = \frac{1}{2} \Big( \pm(2J_S+J_D) \pm \sqrt{J_D^2+4\Delta_S^2}\Big) ~~ (2 ~fold ~ degenerate)
\end{split}
\end{equation}

Fig.\ref{gap_strain_dft} shows the gap at the $\Gamma$ point of MBS from DFT vs. the 
strain along the $\hat{z}$-direction.  
This calculation shows that strains is in the range of around $\pm 2\%$ MBS in 
either direction will convert MBS from a Weyl semimetal into a normal insulator.  
An estimate of $\Delta_S$ and $\Delta_D$ has also been shown.
\ifpdf
\begin{figure}
\centering
\includegraphics[width=0.9 \textwidth]{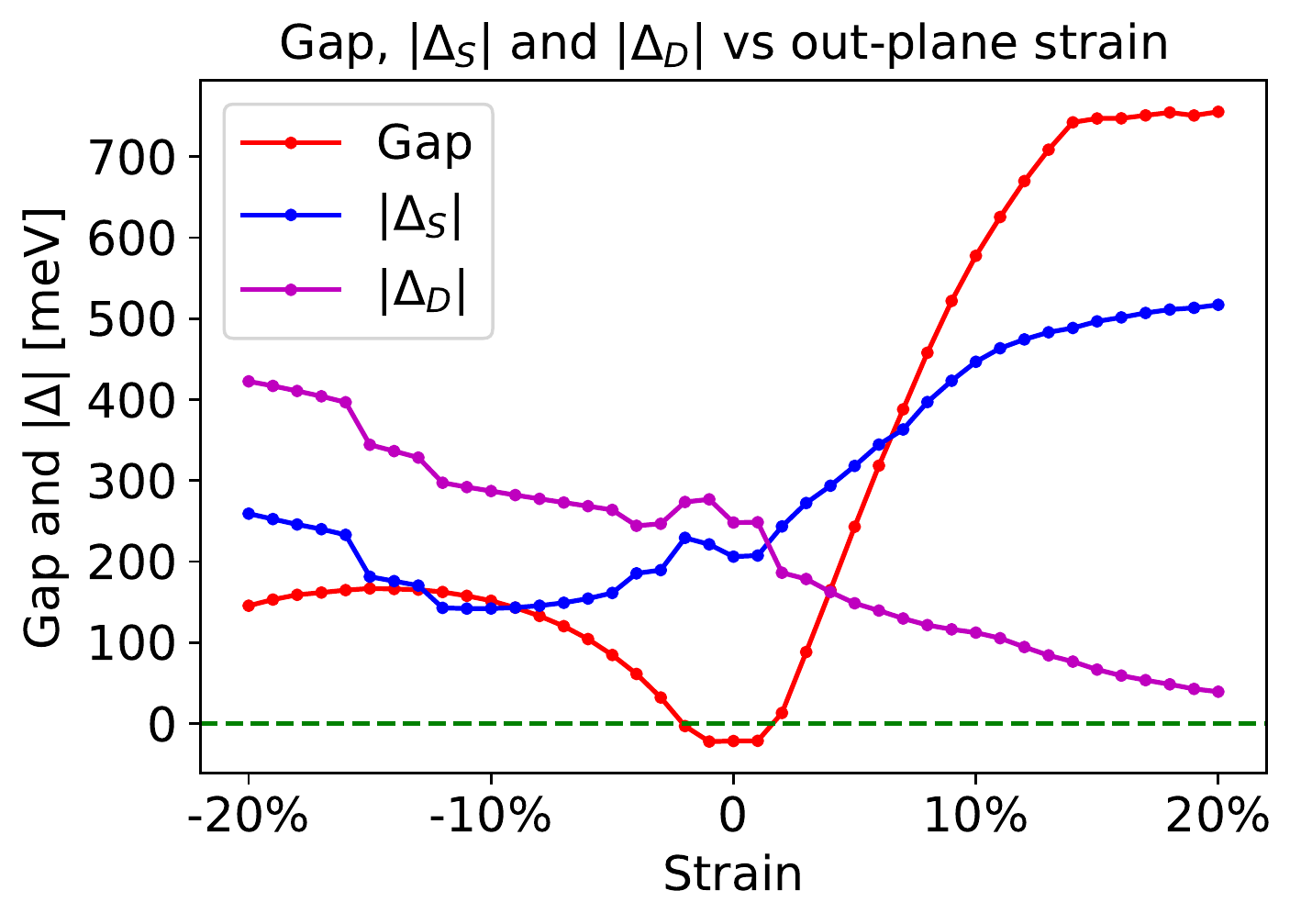}
\caption{Gap at the $\Gamma$ point of MBS and the estimated values of $\Delta_S$ and $\Delta_D$ from DFT {\it vs.} strain along $\hat{z}$-direction. }\label{gap_strain_dft}
\end{figure}
\fi

\subsection*{Weyl semimetal in (M,N) = (1,1) Superlattices}
As shown in Fig. 4 in the maintext, stronger hybridization between surface states within the topological insulator layers trends to favor Weyl semimetal in (M,N) = (1,1) superlattices.
The external magnetic field needed to realize the ferromagnetic state is expected to be dramatically lower in (M,N) = (1,1) superlattices than in a simple (M,N)=(1,0) MTI compound 
because of the non-magnetic spacer between magnetic layers.  
A natural candidate for the magnetized topological insulator layers is MnBi$_2$Te$_4$, which has the weakest same-layer surface hybridization.
although MnBi$_2$Te$_4$/Bi$_2$Te$_3$ has been shown to be an insulator in Fig. 5 in the maintext, we may also consider 
using other non-magnetic topological insulator layers, such as Bi$_2$Se$_3$, Bi$_2$Sb$_3$, or Sb$_2$Te$_3$ as illustrated in Fig. \ref{band_superlattice} (e). 
Since these alternatives all contain one atom with a smaller radius than that in Bi$_2$Te$_3$, a smaller lattice constant is expected, thus also a stronger hybridization between same-layer surface states. 

In Fig. \ref{band_superlattice} (a) the gap of a MnBi$_2$Te$_4$/Bi$_2$Se$_3$ superlattice at the $\Gamma$ point calculated from DFT is plotted {vs.} 
variations of van der Waals spatial gap (illustrated in Fig. \ref{band_superlattice}(b)) for different lateral lattice constant choices.
The red curve retains the lattice constant of MBT, the green curve uses the lattice constant of Bi$_2$Se$_3$ and the blue curve uses
the average lateral lattice constant.  If we keep the lattice constant of MnBi$_2$Te$_4$, the superlattice is a Weyl semimetal. 
The bandstructure is shown in Fig. \ref{band_superlattice} (c).
Note that a localized state appears in the Dirac-cone energy window that is not described by the Dirac cone model. 
These states are localized near the middle of the septuple layers and can accidentally have energies in the low-energy sector.  
Their position is sensitive to the on-site electron-electron interaction Mn atom +U parameter in the DFT calculations 
as shown in Fig. \ref{band_superlattice} (d).
The results from DFT confirm the expectation from the coupled Dirac cone model, that the superlattice is a Weyl semimetal when $\Delta_S$ in the 
non-magnetic topological insulators layers is larger than around 105 meV, with the $\Delta_S \approx 84 meV $ in MnBi$_2$Te$_4$.

\ifpdf
\begin{figure}
\centering
\includegraphics[width=0.9 \textwidth]{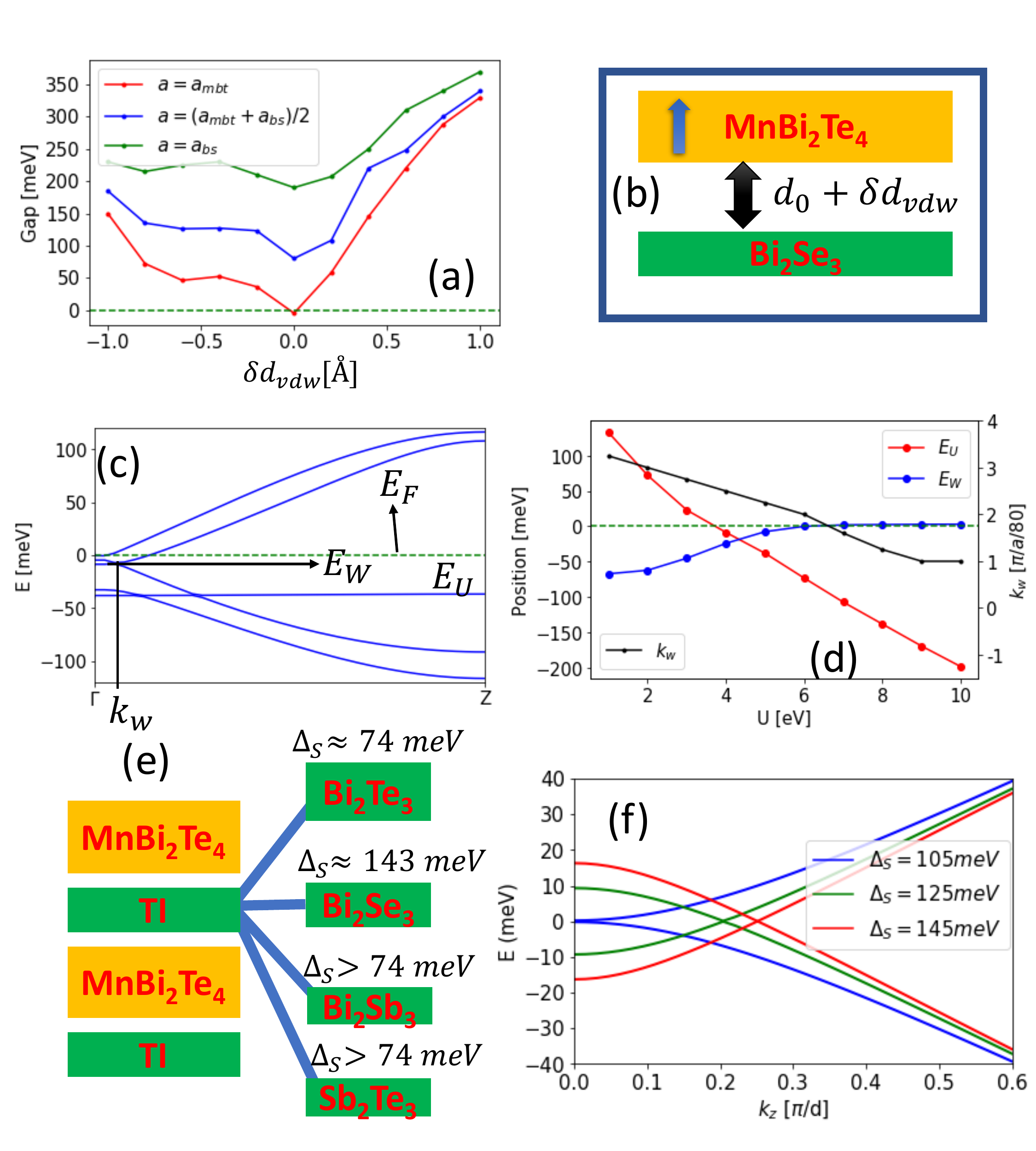}
\caption{ Electronic structure of superlattices with alternating septuple MBT and quintuple topological insulator(TI) layers.
(a) The gap of MnBi$_2$Te$_4$/Bi$_2$Se$_3$ superlattices at the $\Gamma$ point {vs.} the van der Waals spatial gap (b) at 
different lateral lattice constants: the red curve uses the lattice constant of MBT, and the green curve uses the lattice constant of Bi$_2$Se$_3$. 
(c) bandstructure calculated from DFT, which shows that a MnBi$_2$Te$_4$/Bi$_2$Se$_3$ superlattice with the lateral 
lattice constant of MBT is a Weyl semimetal and that a localized subband appears at low energies.
The localized state lies outside of the Dirac-cone model Hilbert space.
The energies of the Weyl point and the localized orbital are plotted as a function of the DFT +U Mn atom interaction parameter in (d).
(f) illustrates Dirac cone model bandstructures calculated for MTI/TI superlattices with $\Delta_S$ parameters in the TI layers that model the 
topological insulator layers illustrated in (e) and the $\Delta_S$ parameter in the MTI layer fixed at 84 meV, corresponding to MBT
}\label{band_superlattice}
\end{figure}
\fi

\end{document}